\RequirePackage{fix-cm}
\documentclass[smallextended]{svjour3}       
\smartqed  
\usepackage[utf8x]{inputenc}

\usepackage{alltt}
\newcommand{\negskip}{\vspace*{-2ex}}
\newcommand{\ps}{$\psi$}
\newcommand{\ph}{$\phi$}
\newcommand{\rh}{$\rho$}
\newcommand{\al}{$\alpha$}
\newcommand{\var}{$\varphi$}
\newcommand{\be}{$\beta$}
\newcommand{\bs}{$\backslash$}

\newenvironment{myverbatim}
{\vspace*{-1ex}\begin{alltt}\small}
	{\end{alltt}\vspace*{-1ex}}
\usepackage{verbatim}
\usepackage{qcircuit}
\usepackage{subfigure}
\usepackage{amsmath}
\usepackage{amssymb}
\usepackage{braket}
\usepackage{color}
\usepackage{url}
\usepackage{times}
\usepackage{ulem} 
\renewcommand{\>}{\rangle}
\newcommand{\<}{\langle}
\newcommand{\sdot}{\cdot}
\newcommand{\zero}{\textbf{0}}
\def\h{\ensuremath{\mathcal{H}}}
\def\lh{\ensuremath{\mathcal{L(H)}}}

\newcommand{\tr}{{\rm tr}}
\newcommand{\law}{{\bf L}}

\begin{document}
	
	\title{Symbolic Reasoning about Quantum Circuits in Coq
																		  
	}
	
	
	\author{Wenjun Shi \and
		Qinxiang Cao \and \\
		Yuxin Deng  \and 
		Hanru Jiang \and
		Yuan Feng
	}
	\authorrunning{W. Shi et al.}
	
	\institute{East China Normal University, China 
		\and
		Shanghai Jiao Tong University, China \and
		East China Normal University, China  \and
		Yanqi Lake Beijing Institute of Mathematical Sciences and Applications, China \and
		University of Technology Sydney, Australia	
	}
	
	\date{This is a pre-print of an article published in Journal of Computer Science and Technology. \\
	The final authenticated version is available online at: https://doi.org/10.1007/s11390-021-1637-9}

	\maketitle
	
	\begin{abstract}
		A quantum circuit is a computational unit that transforms an input quantum state to an output state. A natural way to reason about its behavior is to compute explicitly the unitary matrix implemented by it. However, when the number of qubits increases, the matrix dimension grows exponentially and the computation becomes intractable.
		
		In this paper, we propose a symbolic approach to reasoning about quantum circuits. It is based on a small set of laws involving some basic manipulations on vectors and matrices. This symbolic reasoning scales better than the explicit one and is well suited to be automated in Coq, as demonstrated with some typical examples.
	\end{abstract}

	\section{Introduction}
	A quantum circuit is a natural model of quantum computation~\cite{NC11}. It is a computational unit that transforms an input quantum state to an output state. 
	Once a quantum circuit is designed to implement an algorithm, it is indispensable to analyze the circuit and ensure that it indeed conforms to the requirements of the algorithm. When a large number of qubits are involved, manually reasoning about a circuit's behavior is tedious and error-prone.
	A way of reasoning about quantum circuits (semi-) automatically and reliably is to mechanize the reasoning procedure in an interactive theorem prover, such as the Coq proof assistant~\cite{Coq}. For example, Paykin et al.~\cite{PRZ17} verified a few quantum algorithms in Coq, using some semi-automated strategies to generate machine-checkable proofs.
	
	Existing approaches have apparent drawbacks in both efficiency and human readability. Quantum states and operations are represented and computed using matrices explicitly, and their comparison is made in an element-wise way, thus is highly non-scalable with respect to qubit numbers. Furthermore, as the system dimension grows,
	it is almost impossible for human beings to read the matrices printed by the theorem prover.
	
	In this paper, we propose a symbolic approach to reasoning about the behavior of quantum circuits in Coq, which improves both efficiency of the reasoning procedure and readability of matrix representations. The main contributions of this paper include:
	
	\begin{itemize}
		\item 
		A matrix representation in Coq using the Dirac notation~\cite{Dir}, which is commonly used in quantum mechanics. 
		Matrices are represented as combinations of $\ket{0}$, $\ket{1}$, scalars, and a set of basic operators such as tensor product and matrix multiplication.
		Here $\ket{0}$ and $\ket{1}$ are the Dirac notation for 2-dimensional column vectors
		$[1\ 0]^T$ and $[0\ 1]^T$, respectively.
		In this way, we have a concise representation for sparse matrices, commonly used in quantum computation.
		\item
		A tactic library for (semi-)automated symbolic reasoning about matrices.
		The tactics are based on a small set of inference laws (lemmas in Coq).
		The key idea is to reduce matrix multiplications in the form of $\braket{i|j}$ into scalars, and simplify the matrix representation by absorbing ones and eliminating zeros.
		In this way, our approach reasons about matrices by (semi-)automated rewriting instead of actually computing matrices, and outperforms Paykin et al.'s computational approach~\cite{PRZ17}, as shown in proving the functional correctness of some typical quantum algorithms in Section~\ref{sec:casestudy}.
	\end{itemize}
	
	We illustrate the intuition of our tactics by the following simple example which computes the result of applying the Pauli-X gate to the $\ket{0}$ state. 
	In an explicit matrix-vector multiplication form, it reads as follows.
	$$
	X\ket{0} = 
	\begin{bmatrix} 0 & 1 \\ 1 & 0 \end{bmatrix}
	\begin{bmatrix} 1 \\ 0 \end{bmatrix} = 
	\begin{bmatrix} 
	0 \times 1 + 1 \times 0 \\ 
	1 \times 1 + 0 \times 0
	\end{bmatrix} =
	\begin{bmatrix} 0 \\ 1 \end{bmatrix} =
	\ket{1}
	$$
	and four multiplications are required for the whole computation. 
	By contrast, if we use the Dirac notation for $X$ and apply distribution and associativity laws, then
	$$
	\begin{array}{rcl}
	X\ket{0} & = &  \left(\ket{0}\bra{1} + \ket{1}\bra{0}\right)\ket{0} \\
	& = &  \ket{0}\braket{1|0} + \ket{1}\braket{0|0} \\
	& = &  0 \ket{0} + 1 \ket{1} \\
	& = & \ket{1}.
	\end{array}
	$$
	Note that 
	the two terms $\braket{1|0}$ and $\braket{0|0}$  are reduced (symbolically) to 0 and 1, respectively. Consequently, no multiplication is required at all.
	
	The rest of the paper is structured as follows. In Section~\ref{sec:pre} we recall some basic notation from linear algebra and quantum mechanics. In Section~\ref{sec:symbolic} we introduce a symbolic approach to reasoning about quantum circuits. In Section~\ref{sec:prob} we discuss some problems in representing matrices using Coq's type system and our solutions. 
	In Section~\ref{sec:cir} we propose two notions of equivalence for quantum circuits.
	In Section~\ref{sec:casestudy} we conduct a few case studies. In Section~\ref{sec:related} we discuss some related work. Finally, we conclude the paper in Section~\ref{sec:concl}.
	
	The Coq scripts of our tactic library and the examples used in our case studies are available at the following link
	
	\centerline{\url{https://github.com/Vickyswj/DiracRepr}.}
										 
	\section{Preliminaries}\label{sec:pre}
	For the convenience of the reader, we briefly recall some basic notions
	from linear algebra and quantum theory which are needed in this paper. 
	For more details, we refer the reader to~\cite{NC11}.
	
	\subsection{Basic linear algebra}
	A {\it Hilbert space} $\h$ is a complete vector space equipped with an inner
	product $$\langle\cdot|\cdot\rangle:\h\times \h\rightarrow \mathbb{C}$$
	such that 
	\begin{enumerate}
		\item
		$\langle\psi|\psi\rangle\geq 0$ for any $|\psi\>\in\h$, with
		equality if and only if $|\psi\rangle =0$;
		\item
		$\langle\phi|\psi\rangle=\langle\psi|\phi\rangle^{\ast}$;
		\item
		$\langle\phi|\sum_i c_i|\psi_i\rangle=
		\sum_i c_i\langle\phi|\psi_i\rangle$,
	\end{enumerate}
	where $\mathbb{C}$ is the set of complex numbers, and for each
	$c\in \mathbb{C}$, $c^{\ast}$ stands for the complex
	conjugate of $c$. 
	A vector $|\psi\rangle\in\h$ is {\it normalised} if its length $\sqrt{\langle\psi|\psi\rangle}$ is equal to $1$.
	Two vectors $|\psi\>$ and $|\phi\>$ are
	{\it orthogonal} if $\<\psi|\phi\>=0$. An {\it orthonormal basis} of a Hilbert
	space $\h$ is a basis $\{|i\rangle\}$ where each $|i\>$ is
	normalised and any pair of them is orthogonal.
	
	Let $\lh$ be a set of linear operators on $\h$. For any $A\in
	\lh$, $A$ is {\it Hermitian} if $A^\dag=A$ where
	$A^\dag$ is the adjoint operator of $A$ such that
	$\<\psi|A^\dag|\phi\>=\<\phi|A|\psi\>^*$ for any
	$|\psi\>,|\phi\>\in\h$.
	The fundamental {\it spectral theorem} states that
	the set of all normalised eigenvectors of a Hermitian operator in
	$\lh$ constitutes an orthonormal basis for $\h$. That is, there exists
	a so-called spectral decomposition for each Hermitian $A$ such that
	$$A~=~\sum_i\lambda_i |i\>\<i|,$$
	where the set $\{|i\>\}$ constitutes an orthonormal basis of $\h$, $\{\lambda_i\}$ denotes the set of eigenvalues of $A$, and $|i\>\<i|$ is the projector to
	the corresponding eigenspace of $\lambda_i$.
	A linear operator $A\in \lh$ is {\it unitary} if $A^\dag A=A A^\dag=I_\h$ where $I_\h$ is the
	identity operator on $\h$. 
	The {\it  trace} of $A$ is defined as $\tr(A)=\sum_i \<i|A|i\>$ for some
	given orthonormal basis $\{|i\>\}$ of $\h$. It is worth noting that the
	trace function is actually independent of the orthonormal basis
	selected. It is also easy to check that the trace function is linear and
	$\tr(AB)=\tr(BA)$ for any  $A,B\in \lh$.
	
	Let $\h_1$ and $\h_2$ be two Hilbert spaces. Their {\it tensor product} $\h_1\otimes \h_2$ is
	defined as a vector space consisting of
	linear combinations of the vectors
	$|\psi_1\psi_2\rangle=|\psi_1\>|\psi_2\rangle =|\psi_1\>\otimes
	|\psi_2\>$ with $|\psi_1\rangle\in \h_1$ and $|\psi_2\rangle\in
	\h_2$. Here the tensor product of two vectors is defined by a new
	vector such that
	$$\left(\sum_i \lambda_i |\psi_i\>\right)\otimes
	\left(\sum_j\mu_j|\phi_j\>\right)=\sum_{i,j} \lambda_i\mu_j
	|\psi_i\>\otimes |\phi_j\>.$$ Then $\h_1\otimes \h_2$ is also a
	Hilbert space where the inner product is defined in the following way:
	for any $|\psi_1\>,|\phi_1\>\in\h_1$ and $|\psi_2\>,|\phi_2\>\in
	\h_2$,
	$$\<\psi_1\otimes \psi_2|\phi_1\otimes\phi_2\>=\<\psi_1|\phi_1\>_{\h_1}\<
	\psi_2|\phi_2\>_{\h_2},$$ where $\<\cdot|\cdot\>_{\h_i}$ is the inner
	product of $\h_i$. For any $A_1\in \mathcal{L}(\h_1)$ and $A_2\in
	\mathcal{L}(\h_2)$, $A_1\otimes A_2$ is defined as a linear operator
	in $\mathcal{L}(\h_1 \otimes \h_2)$ such that for each
	$|\psi_1\rangle \in \h_1$ and $|\psi_2\rangle \in \h_2$,
	$$(A_1\otimes A_2)|\psi_1\psi_2\rangle = A_1|\psi_1\rangle\otimes
	A_2|\psi_2\rangle.$$

	\subsection{Basic quantum mechanics}\label{sec:bqm}
	According to von Neumann's formalism of quantum mechanics~\cite{vN55}, an isolated physical system is associated with a
	Hilbert space which is called the {\it state space} of the system. A {\it pure state} of a
	quantum system is a normalised vector in its state space, and a
	{\it mixed state} is represented by a density operator on the state
	space. Here a density operator $\rho$ on Hilbert space $\h$ is a
	positive linear operator such that $\tr(\rho)= 1$. 
	Another
	equivalent representation of a density operator is an 
	ensemble~\cite{NC11} of pure states. In particular, given an ensemble
	$\{(p_i,|\psi_i\rangle)\}$ where $p_i \geq 0$, $\sum_{i}p_i=1$,
	and $|\psi_i\rangle$ are pure states, then
	$\rho=\sum_{i}p_i|\psi_i\>\langle\psi_i|$ is a density
	operator.  Conversely, each density operator can be generated by an
	ensemble of pure states in this way.  
	Finally, a pure state can be regarded as a special mixed state.
	
	The state space of a composite system (for example, a quantum system
	consisting of many qubits) is the tensor product of the state spaces
	of its components. Note that in general, the state of a
	composite system cannot be decomposed into a tensor product of the
	reduced states on its component systems. A well-known example is the
	2-qubit state
	$$|\Psi\>=\frac{1}{\sqrt{2}}(|00\>+|11\>) .$$
	This kind of state is called an {\it entangled state}.
	Entanglement is an important feature of quantum mechanics
	which has no counterpart in the classical world, and is the key to many
	quantum information processing tasks.
	
	Let $|\psi\>$ be a state vector and $\theta$ a real number.
	In quantum mechanics, the state $e^{i\theta}|\psi\>$ is considered to be equal to $|\psi\>$, up to the global phase factor $e^{i\theta}$. The reason is that from an observational point of view, global phases are irrelevant to the observed properties of the physical system under consideration and can thus be ignored as far as quantum states are concerned~\cite{NC11}. 
	
	The evolution of a closed quantum system is described by a unitary
	operator on its state space. If the states of the system at times
	$t_1$ and $t_2$ are $|\psi_1\>$ and $|\psi_2\>$, respectively, then
	$|\psi_2\>=U |\psi_1\>$ for some unitary operator $U$ which
	depends only on $t_1$ and $t_2$. A convenient way to understand unitary operators is in terms of their matrix representations. In fact, the unitary operator and matrix viewpoints turn out to be completely equivalent. An $m$ by $n$ complex unitary matrix $U$ with entries $U_{ij}$ can be considered as a unitary operator sending vectors in the vector space $\mathbb{C}^n$ to the vector space $\mathbb{C}^m$, under matrix multiplication of the matrix $U$ by a vector in $\mathbb{C}^n$.
	
	We often denote a single qubit as a vector $|\psi\> = \alpha|0\> + \beta|1\>$ parameterized by two complex numbers satisfying the condition $|\alpha|^2+|\beta|^2=1$. A unitary operator for a qubit is then described by a $2\times 2$ unitary matrix. Quantum circuits are a popular  model for quantum computation, where quantum gates usually stand for basic unitary operators whose mathematical meanings are given by appropriate unitary matrices. Some commonly used quantum gates to appear in the current work include the $1$-qubit Hadamard gate $H$, the Pauli gates $I_2, X, Y, Z$, the controlled-NOT gate $CX$ performed on two qubits, and the $3$-qubit Toffoli gate. Their matrix representations are given below:
	\[
	I_2=\left(%
	\begin{array}{cc}
	1 & 0 \\
	0 & 1 \\
	\end{array}%
	\right) ,  \qquad
	X=\left(%
	\begin{array}{cc}
	0 & 1 \\
	1 & 0 \\
	\end{array}%
	\right), \qquad Y=\left(%
	\begin{array}{cc}
	0 & -i \\
	i & 0 \\
	\end{array}%
	\right),  \qquad Z=\left(%
	\begin{array}{cc}
	1 & 0 \\
	0 & -1 \\
	\end{array}%
	\right),
	\]
	\[
	H=\frac{1}{\sqrt{2}}\left(%
	\begin{array}{cc}
	1 & 1 \\
	1 & -1 \\
	\end{array}%
	\right), \qquad  
	CX=\left(%
	\begin{array}{cccc}
	1 & 0 & 0 & 0 \\
	0 & 1 & 0 & 0 \\
	0 & 0 & 0 & 1 \\
	0 & 0 & 1 & 0
	\end{array}%
	\right), \qquad
	TOF=\left(%
	\begin{array}{cccccccc}
	1 & 0 & 0 & 0 & 0 & 0 & 0 & 0\\
	0 & 1 & 0 & 0 & 0 & 0 & 0 & 0\\
	0 & 0 & 1 & 0 & 0 & 0 & 0 & 0 \\
	0 & 0 & 0 & 1 & 0 & 0 & 0 & 0\\
	0 & 0 & 0 & 0 & 1 & 0 & 0 & 0\\
	0 & 0 & 0 & 0 & 0 & 1 & 0 & 0\\
	0 & 0 & 0 & 0 & 0 & 0 & 0 & 1\\
	0 & 0 & 0 & 0 & 0 & 0 & 1 & 0
	\end{array}%
	\right) .
	\]
	
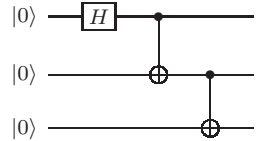
\begin{figure}[htbp]
		\[
		\Qcircuit @C=.8em @R=1.7em {
			\lstick{\ket{0}} & \qw & \gate{H} & \qw & \ctrl{1} & \qw & \qw & \qw & \qw \\
			\lstick{\ket{0}} & \qw & \qw & \qw & \targ & \qw & \ctrl{1} & \qw & \qw  \\
			\lstick{\ket{0}} & \qw & \qw & \qw & \qw & \qw & \targ & \qw & \qw  \\
		}
		\]
		\caption{A circuit for creating the 3-qubit GHZ state}\label{GHZ}
	\end{figure}

	For example, in Figure~\ref{GHZ} we display a circuit that can generate the $3$-qubit Greenberger–Horne–Zeilinger state (GHZ state)~\cite{GHZ}, which is $\frac{|000\>+|111\>}{\sqrt{2}}$. In the circuit, a Hadamard gate is applied on the first qubit, then two controlled-NOT gates are used, with the first qubit controlling the second, which in turn controls the third.
	
	A quantum {\it measurement} is described by a
	collection $\{M_m\}$ of measurement operators, where the indices
	$m$ refer to the measurement outcomes. It is required that the
	measurement operators satisfy the completeness equation
	$\sum_{m}M_m^{\dag}M_m=I_\h$.
	If the state of a quantum system is $|\psi\>$ immediately before the measurement, then the probability that result $m$ occurs is given by
	\[p(m)=\<\psi| M_m^\dag M_m|\psi\> ,\]
	and the state of the system after the measurement is
	$\frac{M_m |\psi\>}{\sqrt{p(m)}} .$
	If the states of the system at times
	$t_1$ and $t_2$ are mixed, say $\rho_1$ and $\rho_2$, respectively, then
	$\rho_2=U\rho_1U^{\dag}$ after the unitary operation $U$ is applied on the system.
	For the same measurement $\{M_m\}$ as above, if the system is in the mixed state $\rho$, then the probability
	that measurement result $m$ occurs is given by
	$$p(m)=\tr(M_m^{\dag}M_m\rho),$$ and the state of the post-measurement system
	is $\frac{M_m\rho M_m^{\dag}}{p(m)}$ provided that $p(m)>0$.

	\section{Symbolic reasoning}\label{sec:symbolic}
	\label{sec:sym}
	In this section, we first introduce the terms that will appear in our symbolic reasoning and some laws that  are useful for reducing terms. Then we present in detail the strategeis that we developed to simplify circuits.
	
	\begin{table}[t]
		\begin{center}
			\begin{tabular}{ll}
				\hline 
				Scalars: & $\mathbb{C}$ \\
				Basic vectors:  & $|0\>$, $|1\>$ \\ 
				Operators: & $\sdot$, $\times$, $+$, $\otimes$, $\dag$\\
				\hline
				Laws: & \law 1\quad $\<0|0\> = \<1|1\> = 1$,\ \ $\<0|1\> = \<1|0\> = 0$ \\
				& \law 2\quad Associativity of $\sdot,\ \times,\ +,\ \otimes$\\
				& \law 3\quad  $0 \sdot A_{m\times n} = \zero_{m\times n}$,\ \ $c\sdot \zero=\zero$,\ \ $1 \sdot A = A$\\
				& \law 4\quad  $c \sdot (A + B) = c \sdot A + c \sdot B$\\
				& \law 5\quad $c \sdot (A \times B) = (c \sdot A)\times B = A \times (c \sdot B)$ \\
				& \law 6\quad $c \sdot (A \otimes B) = (c \sdot A)\otimes B = A \otimes (c \sdot B)$ \\
				& \law 7\quad $\zero_{m\times n} \times A_{n\times p} = \zero_{m\times p} = A_{m\times n} \times \zero_{n\times p}$\\ 
				& \law 8\quad $I_m \times A_{m\times n} =A_{m\times n} = A_{m\times n}\times I_n $, \ \ $I_m\otimes I_n = I_{mn}$\\
				& \law 9\quad $\zero + A = A = A + \zero$\\
				& \law 10\quad $\zero_{m\times n} \otimes A_{p\times q} = \zero_{mp\times nq} = A_{p\times q}\otimes \zero_{m\times n}$\\
				& \law 11\quad $(A + B) \times C = A \times C + B\times C$,\ \ $C\times (A + B) = C \times A + C \times B$\\
				&  \law 12\quad $(A + B) \otimes C = A \otimes C + B\otimes C$,\ \ $C\otimes (A + B) = C \otimes A + C \otimes B$\\
				& \law 13\quad $(A\otimes B)\times (C\otimes D) = (A\times C)\otimes (B \times D)$\\
				& \law 14\quad $(c \sdot A)^\dag = c^* \sdot A^\dag$,\ \ $(A \times B)^\dag = B^\dag \times A^\dag$\\ 
				& \law 15\quad $(A + B)^\dag = A^\dag + B^\dag$,\ \ $(A \otimes B)^\dag = A^\dag \otimes B^\dag$ \\ 
				& \law 16\quad $(A^\dag)^\dag = A$\\
				\hline
			\end{tabular}
		\end{center}
		\caption{Terms and laws}\label{t:core}
	\end{table}
	
	\subsection{Terms and laws}
	Our symbolic reasoning is based on terms constructed from scalars and basic vectors using some constructors:
	
	\begin{itemize}
		\item 
		Scalars are complex numbers. We write $\mathbb{C}$ for the set of complex numbers. Our formal treatment of complex numbers is based on the definitions and lemmas in the library of Paykin et al.~\cite{PRZ17} for complex numbers, which, in turn, is adapted from the Coquelicot library~\cite{BLM}.
		\item 
		Basic vectors are the base states of a qubit, i.e., $|0\>$ and $|1\>$ in the Dirac notation.
		Mathematically, $|0\>$ stands for the vector $[1\ 0]^T$ and $|1\>$ for $[0\ 1]^T$.
		\item 
		Constructors include the scalar product $\sdot$, matrix product $\times$, matrix addition $+$, tensor product $\otimes$, and the conjugate transpose $A^\dag$ of a matrix $A$.
	\end{itemize}
	
	In Dirac notations, $\<0|$ represents the dual of $|0\>$, i.e. $|0\>^\dag$; similarly for $\<1|$. The term $\<j|\times |k\>$ is abbreviated into $\<j|k\>$, for any $j,k\in\{0,1\}$. This notation introduces an intuitive explanation of quantum operation. For example, the effect of the $X$ operator is to map
	$ |0\>$ into $|1\>$ and  $|1\>$ into  $|0\>$.
	Thus we define $X$ in Coq as $|0\>\<1| + |1\>\<0|$, instead of $\begin{bmatrix} 0\, 1\\ 1\, 0\end{bmatrix}$.
	Then it is obvious that
	$X |0\> = |1\>\<0|0\> + |0\>\<1|0\> = |1\>$ and similarly for $X|1\>$.
	
	
	Some commonly used vectors and gates can be derived from the basic terms. For example, we define the vectors $|+\>$ and $|-\>$ as follows.
	\[|+\> ~=~  \frac{1}{\sqrt{2}} \sdot |0\> + \frac{1}{\sqrt{2}} \sdot |1\> \qquad\qquad
	|-\> ~ = ~ \frac{1}{\sqrt{2}} \sdot |0\> + (-\frac{1}{\sqrt{2}}) \sdot |1\>\]
	We also define four simple matrices $B_0, ..., B_3$.
	\begin{equation}\label{bj} B_0 ~=~ |0\> \times \<0|\quad \quad
	B_1 ~=~ |0\> \times \<1|\quad \quad
	B_2 ~ =~ |1\> \times \<0| \quad \quad
	B_3 ~=~ |1\> \times \<1|
	\end{equation}
	The Hadamard matrix $H$ is a combination of the four matrices above.
	\[H ~=~ \frac{1}{\sqrt{2}} \sdot B_0 + \frac{1}{\sqrt{2}} \sdot B_1
	+ \frac{1}{\sqrt{2}} \sdot B_2 + (-\frac{1}{\sqrt{2}}) \sdot B_3 \]
	Similarly, the Pauli-X gate and the controlled-NOT gate $CX$ are given as follows. 
	\begin{equation}\label{eq:cx}
		X ~=~ B_1 + B_2 \qquad\qquad
		CX ~=~ B_0 \otimes I_2 + B_3\otimes X
	\end{equation}
	
	Notice that tensor products of matrices from $\{B_j : j= 0,1,2,3\}$ constitute an orthonormal basis for the space of all matrices of dimension $2^n$, for all $n\geq 1$. Thus any matrix that appeared in the computation of a quantum circuit can be represented as a term.
	
	
	Suppose the state of a quantum system is represented by a vector. The central idea of our symbolic reasoning is to employ the laws in Table~\ref{t:core} to rewrite terms, trying to put together the basic vectors and simplify them using the laws in {\law}1. Technically, we design a series of strategies for that purpose.
	%
	%
	
	\subsection{Soundness of the laws}\label{sec:soundness}
	Let $T_1 = T_2$ be a law that relates two terms $T_1$ and $T_2$. We say that the law is sound if both terms can be reduced to exactly the same matrix, after their metavariables (e.g., $\ket{0}$) are instantiated by concrete matrices (e.g., $[1\ 0]^T$).
	We define a strategy called {\tt orthogonal$\_$reduce} to verify that the laws in {\law}1 are sound. In this case, we  explicitly represent $|0\>$ and $|1\>$ as matrices. Both of them are $2\times 1$ matrices, so we can use matrix multiplication to prove that the corresponding elements in the matrices on both sides of the equation are the same. For example, the law $\<0|0\> = 1$ is actually shown via $[1\ 0] \begin{bmatrix} 1 \\ 0\end{bmatrix} = 1$. We add the laws in {\law}1 to the set named {\tt S\_db}. 
	The laws in {\law}2-16 are also proved through explicit matrix representation.
	\begin{theorem}
		All the laws in Table~\ref{t:core}  are sound.
	\end{theorem}
	
	We have given a formal soundness proof in Coq for each of the laws in Table~\ref{t:core}. We collect the soundness properties in a library that contains  many useful properties about matrices. 
	
	\subsection{Strategy for basic matrices}
	The four matrices $B_0, ..., B_3$ defined in \eqref{bj} are the basic building blocks to represent $2\times 2$ matrices, and are intensively used in our symbolic reasoning. 
	We design a strategy called {\tt base$\_$reduce} to prove some equations about them acting on the base states $|0\>$ and $|1\>$. %
	For example, let us consider the equation $B_0 \times |0\> = |0\>$. We first represent $B_0$ by $|0\> \times \<0|$, then use the associativity of matrix multiplication to form the subterm $\<0| \times |0\>$. Now we can use the proved laws in {\law}1 to rewrite $\<0| \times |0\>$ into $1$. The last step is to deal with scalar multiplications. We add these equations to the set named {\tt B\_db}.
	
	\subsection{Strategy for the Pauli and Hadamard gates}
	The Pauli and Hadamard gates are probably the most widely used single-qubit gates.
	We introduce a strategy called {\tt gate$\_$reduce} to prove some equations about the matrices $I_2,X,Y,Z,H$ acting on base states. %
	For example, consider the equation $X \times |0\> = |1\>$. We first expand $X$ into $B_1 + B_2$. In order to prove the equation $(B_1 + B_2) \times |0\> = |1\>$, we use the distributivity of matrix multiplication over addition to  rewrite the left hand side of the equation into the sum of $B_1 \times |0\>$ and $B_2 \times |0\>$.
	Then we employ the proved laws in {\tt B\_db} to rewrite them. Eventually, we deal with scalar multiplications and cancel zero matrices. We add these equations to the set named {\tt G\_db}.
	
	Furthermore, we add into {\tt S\_db} and {\tt G\_db} some commonly used equations about the matrices $B_0, ..., B_3$ and $I_2,X,Y,Z,H$ acting on states $|+\>$ and $|-\> $. For example, we have $H|+\> = |0\>$ and $H |-\> = |1\>$, etc.
	
	\subsection{Strategy for circuits}\label{sec:circ}
	We propose a strategy called {\tt operate$\_$reduce} that puts together all the 
	results above to reason about circuits applied to input states represented in the vector form. %
	We will have a close look at the strategy by using an example.
	Let us revisit the 3-qubit GHZ state. The
	state can be generated by applying the circuit in Figure~\ref{GHZ} to the initial state $|0\> \otimes |0\> \otimes |0\>$. We would like to verify that the output state is indeed what we expect by establishing the following equation.
	\begin{equation}\label{e:i}\begin{array}{rl}
	& (I_2 \otimes CX) \times (CX \otimes I_2) \times (H \otimes I_2 \otimes I_2) \times (|0\> \otimes |0\> \otimes |0\>) \\
	=  & \frac{1}{\sqrt{2}} \sdot (|0\> \otimes |0\> \otimes |0\>) + \frac{1}{\sqrt{2}} \sdot (|1\> \otimes |1\> \otimes |1\>)
	\end{array} \end{equation}	
	
	We first use the right associativity of tensor product and matrix multiplication to change the expression on the left hand side in \eqref{e:i} into
	\begin{equation}\label{e:ii}\begin{array}{rl}
	& (I_2 \otimes CX) \times ((CX \otimes I_2) \times ((H \otimes (I_2 \otimes I_2)) \times (|0\> \otimes (|0\> \otimes |0\>)))).
	\end{array} \end{equation}	
    Then we calculate the inner layer matrix multiplications in sequence. We first calculate $(H \otimes (I_2 \otimes I_2)) \times (|0\> \otimes (|0\> \otimes |0\>))$. We exploit the law in {\law}13 to change a matrix product of tensored terms into a tensor product of matrix multiplications. Then we employ the  laws established in {\tt G\_db} and {\tt B\_db} to rewrite terms. This transformation is as follows.
	\[\begin{array}{rl}
	& (H \otimes (I_2 \otimes I_2)) \times (|0\> \otimes (|0\> \otimes |0\>)) \\
	=  & (H \times |0\>) \otimes ((I_2 \times |0\>) \otimes (I_2 \times |0\>)) \\
	=  & |+\> \otimes (|0\> \otimes |0\>) \\
	\end{array} \]
	Correspondingly, the expression in \eqref{e:ii} turns into
	\[\begin{array}{rl}
	& (I_2 \otimes CX) \times ((CX \otimes I_2) \times (|+\> \otimes (|0\> \otimes |0\>))) .
	\end{array} \]

	The inner layer matrix multiplication to be calculated next is the following expression: $(CX \otimes I_2) \times (|+\> \otimes (|0\> \otimes |0\>))$. Different from the calculation of the previous layer, we have to expand the multiple-qubit quantum gates first. Here is the CX gate.
	After this step, we use the distributivity of tensor product and matrix product over addition to rewriting it. We also exploit the law in {\law}13 as well as the equations in {\tt G\_db} and {\tt B\_db}. So the inference goes  as follows.
	\[\begin{array}{rl}
	& (CX \otimes I_2) \times (|+\> \otimes (|0\> \otimes |0\>)) \\
	=  & ((B_0 \otimes I_2 + B_3\otimes X) \otimes I_2) \times (|+\> \otimes (|0\> \otimes |0\>)) \\
	=  & ((B_0 \otimes I_2 \otimes I_2 + B_3\otimes X \otimes I_2) \times (|+\> \otimes (|0\> \otimes |0\>)) \\
	=  & ((B_0 \otimes (I_2 \otimes I_2)) \times (|+\> \otimes (|0\> \otimes |0\>)) + ((B_3\otimes (X \otimes I_2) \times (|+\> \otimes (|0\> \otimes |0\>)) \\
	=  & ((B_0 \times |+\>) \otimes (I_2 \times |0\>) \otimes (I_2 \times |0\>)) + ((B_3 \times |+\>) \otimes (X \times |0\>) \otimes (I_2 \times |0\>)) \\
	=  & \frac{1}{\sqrt{2}} \sdot (|0\> \otimes (|0\> \otimes |0\>)) + \frac{1}{\sqrt{2}} \sdot (|1\> \otimes (|1\> \otimes |0\>))
	\end{array} \]
	Other laws such as the associativity and distributivity of scalar multiplication may also be used in the inference, though they are not demonstrated in this example.
	
	We continue in this way until no more matrix multiplication is possible. It is worth noting that if there is a zero matrix in the summands of a certain layer of calculation, it can be eliminated immediately, without being carried  into the next layer of calculation. Finally, we simplify the expressions about complex numbers.
	
	The above steps appear a bit complex, but they can be fully automated in Coq, fortunately. The script for implementing  the strategy {\tt operate$\_$reduce} is as follows.
	\begin{myverbatim}
		\negskip
		Ltac inner_reduce :=
		  unfold_operator;
		  kron_plus_distr;
		  isolate_scale;
		  assoc_right;
		  try repeat rewrite Mmult_plus_distr_l;
		  try repeat rewrite Mmult_plus_distr_r;
		  repeat rewrite <- Mscale_kron_dist_r;
		  repeat mult_kron;
		  repeat rewrite Mscale_mult_dist_r;
		  repeat (autorewrite with G_db;
		  repeat cancel_0;
		  repeat rewrite Mscale_kron_dist_r);
		  repeat rewrite <- Mmult_plus_distr_l.
		\negskip
		Ltac operate_reduce :=
		  assoc_right;
		  repeat inner_reduce;
		  reduce_scale;
		  unified_base.
		\negskip
	\end{myverbatim}
	
	In summary, using the strategy {\tt operate$\_$reduce}, we can formally prove the equation in \eqref{e:i} automatically.
	
	As we can see, our general framework is to formalize circuits symbolically as terms, and simplify terms involving matrices by (semi-)automated term rewriting instead of actually computing the matrices. The strategies in Sections~\ref{sec:soundness}-\ref{sec:circ} are designed with the common goal: to reduce matrix multiplications in the form of $\braket{i|j}$ into scalars, and simplify the matrix representation by absorbing ones and eliminating zeros.
	This symbolic approach of reasoning about circuits turns out to be effective; see Section~\ref{sec:exp} for more detailed discussion.
	
	
	\subsection{Density matrices as quantum states}\label{sec:density}
	Although it is very intuitive to represent pure quantum states by vectors, there is an inconvenience. In quantum mechanics, the global phase of a qubit is often ignored. For example, we would not distinguish  $|\psi\>$ from $e^{i\theta}|\psi\>$ for any $\theta$. However, when written in vector form, $|\psi\>$ and $e^{i\theta}|\psi\>$ may be different because of the coefficient $e^{i\theta}$ present in the latter but not in the former. 
	Therefore, we use the symbol $\approx$ to denote such an equivalence, i.e. $e^{i\theta}|\psi\> \approx |\psi\>$. As a matter of fact, we can be more general and define an observational equivalence for matrices, as given below.
	\begin{myverbatim}
		\negskip
		Definition obs_equiv \{m n : nat\} (A B : Matrix m n) : Prop :=
   exists c : C, Cmod c = R1 /\bs  c .* A = B.
		\negskip
		Infix "≈" := obs_equiv.
		\negskip
	\end{myverbatim}
	In the above definition, the condition {\tt Cmod c = R1} means that the norm of the complex number {\tt c} is 1 and {\tt c .* A = B} says that the matrix {\tt A} is equal to {\tt B} after a scalar product with the coefficient {\tt c}.
	See Section~\ref{sec:deu} for more concrete examples that use the relation $\approx$. 
	
	Note that if quantum states are represented by density matrices, we have
	\[ (e^{i\theta}|\psi\>)(e^{i\theta}|\psi\>)^\dag ~=~ 
	(e^{i\theta}|\psi\>)(e^{-i\theta}\<\psi|) ~=~
	|\psi\>\<\psi| .\]
	Therefore, the discrepancy entailed by the global phase  disappears: the two vectors $e^{i\theta}|\psi\>$ and $|\psi\>$ correspond to the same density matrix $|\psi\>\<\psi| $.
	Representing states by density matrices can thus omit unnecessary details and in some cases simplify our reasoning. This 
	is a small but useful trick in formal verification of
	quantum circuits, which does not seem to have been exploited in the literature. In Sections~\ref{sec:tele} and~\ref{sec:simon}, we give two examples where the input and output quantum states of the circuits are given in terms of density matrices.
	
	If the state of a quantum system is represented by a density matrix, the reasoning strategies discussed above can still be used. For instance, suppose a system is in the initial state given by density matrix $\rho$. After the execution of a quantum circuit implementing some unitary transformation $U$, the system changes into the new state $\rho'=U\rho U^\dag$. 
	Let $\rho=\sum_j \lambda_j|j\>\<j|$ be its spectral decomposition, where $\lambda_j$ are eigenvalues of $\rho$ and the vectors $|j\>$ 
	the corresponding eigenvectors. It follows that
	\begin{equation}\label{e:iv}\rho' ~=~ U(\sum_j \lambda_j|j\>\<j|)U^\dag ~=~ \sum_j\lambda_j U|j\> (U|j\>)^\dag\ . \end{equation}
	Therefore, we can first simplify $U|j\>$ into a vector, take its dual and then obtain $\rho'$ easily. Our approach to symbolic reasoning also applies in this setting.
	
	We define two functions  {\tt density} and {\tt super} in advance. The former converts states in the vector form into corresponding states in the density matrix form. The latter formalizes the transformation process between states in the density matrix form. 
	\begin{myverbatim}
	    \negskip
		Definition density \{n\} (\ps : Matrix n 1) : Matrix n n := \ps × \ps†.
		Definition super \{m n\} (M : Matrix m n) : Matrix n n -> Matrix m m := 
   fun \rh => M × \rh × M†.
        \negskip
	\end{myverbatim}
	We introduce the simplification strategy called {\tt super$\_$reduce} for states in the density matrix form. 
	\begin{myverbatim}
		\negskip
		Ltac super_reduce:=
		  unfold super,density;                
		(* Expand super and density *)
		  match goal with                      
		(* Match the pattern of target 
		   with U × \ps × \ps† × U† *)
		   ||-context [ (?A × ?B) × ?A† ] =>       
		     match B with
	 	    | (?C × ?C†) =>
		         transitivity ((A × C) × (C† × A†)       
		(* Cast uniform types *)
		     end
		  end;
		  [repeat rewrite <- Mmult_assoc; reflexivity|..];
		  rewrite <- Mmult_adjoint;                    
		(* Extract adjoint *)
		  let Hs := fresh "Hs" in                       
		    match goal with                          
		    ||-context [ (?A × ?B) × ?C† ) = ?D × ?D†]=>
		      match C with
		      | ?A × ?B=> assert (A × B = D) as Hs
		      end                                       
		    end;
		(* Use operate_reduce to prove vector states
		   and rewrite it in density matrix form *)                               
		  [try reflexivity; try operate_reduce |         
		   repeat rewrite Hs; reflexivity]. 
		\negskip
	\end{myverbatim}
	In the above strategy, we first expand the {\tt density} and {\tt super} functions in the target. Next, we match the pattern of the target to see whether it is in the form $U \times |\psi\>  \times \<\psi| \times U^\dagger$ (the middle of the equation in \eqref{e:iv}) and cast uniform types, for the reasons to be discussed in Section~\ref{sec:prob}. Then we exploit the law in {\law}14 to extract adjoint of multiplication terms so the target becomes $U \times |\psi\>  \times (U \times |\psi\>)^\dagger$ as in the right hand side of the equation in \eqref{e:iv}. Finally, we use the strategy {\tt operate$\_$reduce} to conduct the proof for states in vector form and rewrite it back in density matrix form.
	Note that we omit dimensions for presentation purpose, in practice we need to specify these implicit arguments.
	
	\subsection{Mixed state}\label{sec:mix}	
	As mentioned in Section~\ref{sec:bqm}, the mixed state $\rho=\sum_i p_i|\psi_i\>\<\psi_i|$ is an ensemble $\{(p_i,|\psi_i\rangle)\}$ of pure states $|\psi_i\rangle$, where $p_i$ is the probability of the pure state $|\psi_i\>$ with $\sum_{i}p_i=1$. In Coq, we use a list of pairs of real numbers and density operators to define mixed states.
																							 
	\begin{myverbatim}
	    \negskip
		Definition Pure n := (R * (Matrix n n)).
		Definition Mix n := (list (Pure n)).
		\negskip
	\end{myverbatim}

	 Recall that the application of unitary operator $U$ on the mixed state $\rho$ is in the following form:
	 \begin{equation}\label{e:mu}\rho = \sum_i p_i|\psi_i\>\<\psi_i|
	 ~\xrightarrow{U}~ U \rho U^\dag = \sum_ip_j U|\psi_i\> \<\psi_i| U^\dag . 
	 \end{equation}
	 We formalize \eqref{e:mu} in Coq as follows.
	\begin{myverbatim}
		\negskip				 
		Fixpoint UnitMix \{n\} (A : Matrix n n) (m : Mix n): Mix n :=
		  match m with
		  | [] => []
		  | a :: b => (match a with
		              |(x , y) => (x , super A y)
		              end) :: (UnitMix A b)
		  end.
	    \negskip
	\end{myverbatim}
	
	After a measurement, results can be expressed with mixed states, as introduced in Section~\ref{sec:bqm}. For simplicity, we  use projection measurements, so for each measurement operator $M_m$ we have $M_m^{\dag}M_m=M_m$.
	We first define the measurement operators of any dimension as follows, where the two parameters $n$ and $k$ stand for the space dimension  and the position of an active qubit.
	\begin{myverbatim}
	    \negskip
		Definition Mea0 (n k : N) := (I (2^k) ⊗ ∣0⟩⟨0∣ ⊗ I (2^(n-k))).
		Definition Mea1 (n k : N) := (I (2^k) ⊗ ∣1⟩⟨1∣ ⊗ I (2^(n-k))).
		Definition Mea (n k : N) := Mea0 n k .+ Mea1 n k.
	    \negskip
	\end{myverbatim}	
	Then we formalize measurements on the mixed state in Coq as follows. Note that a pure state $\rho$ can be regarded as a special mixed state $[(1,\rho)]$.
	\begin{myverbatim}
	    \negskip
		Fixpoint MeaMix \{n\} (m k : N) (l : Mix n) : Mix n :=
		  match l with
		  | [] => []
		  | a :: b => match a with
		              | (x , y) =>  
		  [((x * (trace((Mea0 m k) × y))), /(trace ((Mea0 m k)× y)).* super(Mea0 m k) y);
		  ((x * (trace((Mea1 m k) × y))), /(trace ((Mea1 m k)× y)).* super(Mea1 m k) y)]
		              end ++ (MeaMix m k b)
		  end.
		  \negskip
	\end{myverbatim}

	\section{Equivalences of circuits}\label{sec:cir}
	In order to judge whether two circuits have the same behavior, we need to formally define reasonable notions of equivalence for circuits in the first place. In this section, we propose two candidate relations: one is called matrix equivalence and the other observational equivalence.
	\subsection{Matrix equivalence}
	\label{sec:mat}
	A natural way of interpreting a quantum circuit without measurements is to consider each quantum gate as a unitary matrix and the whole circuit as a composition of matrices that eventually reduces to a single matrix. From this viewpoint, two circuits are equivalent if they denote the same unitary matrix, that is, matrix equivalence $=$ suffices to stand for circuit equivalence.
	
	\begin{figure}
		\begin{tabular}{rclrcl}
			$X X$ & = & $I_2$ & \hspace{2cm}$\frac{1}{\sqrt{2}} \sdot (X + Z)$ & = &  $H$ \\
			$Y Y$ & = & $I_2$ & $H_2 × CX × H_2$ & = &  $CZ$ \\
			$Z Z$ & = & $I_2$ & $CX \times X_1 \times CX$ & = & $X_1 \times X_2$ \\
			$H H$ & = & $I_2$ & $CX \times Y_1 \times CX$ & = & $Y_1 \times X_2$ \\
			$CX \times CX$ & = & $I_4$ & $CX \times Z_1 \times CX$ & = & $Z_1$\\
			$ HXH$ & = & $Z$ & $CX \times X_2 \times CX$ & = & $X_2$\\
			$ HYH$ & = & $-Y$ & $CX \times Y_2 \times CX$ & = & $Z_1 \times Y_2$\\
			$HZH$ & = & $X$ & $CX \times Z_2 \times CX$ & = & $Z_1 \times Z_2$ 
		\end{tabular}
		\caption{More laws}\label{fig:laws}
	\end{figure}
	
	Directly showing that two matrices are equivalent requires to inspect their elements and compare them component-wisely. Instead, we can take a functional view of matrix equivalence. Let $A, B$ be two $2^m\times 2^m$ matrices, then 
	$A= B$ if and only if $A|v\> = B|v\>$ for any vector $|v\>\in\h$, where $\h$ is the space of all $m$-qubit states.
	\begin{myverbatim}
		\negskip
		Lemma MatrixEquiv_spec: forall \{n\} (A B: Matrix n n),
   A = B <-> (forall v: Vector n, A × v = B × v).
		\negskip
	\end{myverbatim}
	At first sight, it appears difficult to verify whether
	$A|v\> = B|v\>$ for all  vectors $|v\>$, since there are infinitely many vectors in the state space $\h$. However, both matrices $A$ and $B$ represent linear operators, which means that it suffices to consider the vectors in an orthonormal basis of $\h$, where there are only $2^m$ vectors.
	
	In Figure~\ref{fig:laws} we list some laws that are often useful in simplifying circuits before showing that they are equivalent.
	Let us verify the validity of the laws. 
	Take the first one  as an example. Its validity is stated in Lemma {\tt unit$\_$X}. In order to prove that lemma, we apply {\tt MatrixEquiv$\_$spec}  and reduce it to  Lemma {\tt unit$\_$X'}, which can be easily proved by the strategy {\tt operate$\_$reduce}.
	\begin{myverbatim}
		\negskip
		Lemma unit_X : X × X = I_2.
		\negskip
		Lemma unit_X' : forall v : Vector 2,  X × X × v = I_2 × v.	   
	\end{myverbatim}
	
	In the right column of Figure~\ref{fig:laws},
	the subscripts of $X,Y,Z$ and $H$ indicate on which qubits the quantum gates are applied.  For example, $X_2$ means that the Pauli-X gate is applied on the second qubit. Thus, the operation $Y_1 \times X_2$ actually stands for $(Y \otimes I_2) \times (I_2 \otimes X)$. 
	\begin{figure}
		\subfigure[]{
			\begin{minipage}[t]{0.5\linewidth}
				\[
				\Qcircuit @C=.7em @R=1em {
					& \ctrl{2} & \qw & \targ & \qw & \ctrl{2} & \qw & & & \qw & \qswap & \qw & \qw \\
					& & & & & & & \push{\rule{.3em}{0em}=\rule{.3em}{0em}} & & & \qwx & & \\
					& \targ & \qw & \ctrl{-2} & \qw & \targ & \qw & & & \qw & \qswap \qwx & \qw & \qw\\
				}
				\]				
			\end{minipage}%
		}%
		\subfigure[]{
			\begin{minipage}[t]{0.5\linewidth}
				\[
				\Qcircuit @C=.6em @R=1em {
					& \qw & \ctrlo{2} & \qw & \qw & & & \qw & \gate{X} & \ctrl{2} & \gate{X} & \qw & \qw \\
					& & & & & \push{\rule{.3em}{0em}=\rule{.3em}{0em}} & & & & & \\
					& \qw & \targ & \qw & \qw & & & \qw & \qw & \targ & \qw & \qw & \qw \\
				}
				\]
			\end{minipage}%
		}%
		\\ 
		\subfigure[]{
			\begin{minipage}[t]{0.5\linewidth}
				\[
				\Qcircuit @C=.4em @R=.1em {
					& \ctrl{2} & \qw & & & \qw & \gate{\left[
						\begin{array}{cc}
						1 & 0 \\
						0 & e^{i\alpha} \\
						\end{array}%
						\right]} & \qw \\
					& & & \push{\rule{.3em}{0em}=\rule{.3em}{0em}} & & & & \\
					& \gate{\left[
						\begin{array}{cc}
						e^{i\alpha} & 0 \\
						0 & e^{i\alpha} \\
						\end{array}%
						\right]} & \qw & & & \qw & \qw & \qw\\
				}
				\]	
			\end{minipage}%
		}%
		\subfigure[]{
			\begin{minipage}[t]{0.5\linewidth}
				\[
				\Qcircuit @C=.8em @R=1.7em {
					& \qw & \ctrl{2} & \qw & \qw & & & \qw & \ctrl{1} & \qw & \ctrl{2} & \qw & \qw \\
					& \qw & \targ & \qw & \qw & \push{\rule{.3em}{0em}=\rule{.3em}{0em}} & & \qw & \targ & \qw & \qw & \qw & \qw & \\
					& \qw & \targ & \qw & \qw & & & \qw & \qw & \qw & \targ & \qw & \qw & \\
				}
				\]
			\end{minipage}
		}%
		\caption{Some equivalent circuits} \label{Cir}
	\end{figure}
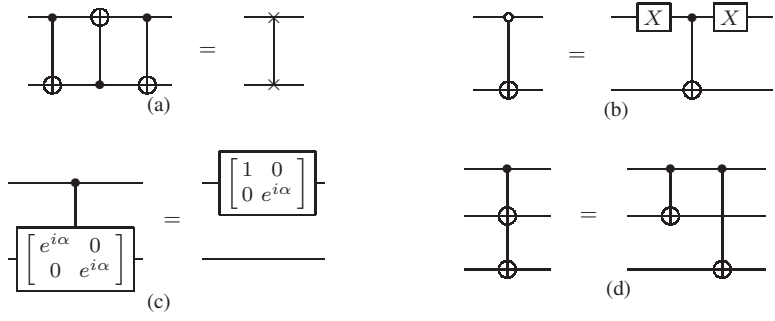
	
	In Figure~\ref{Cir}, we display some equivalent circuits. In diagram (a),  on the right of = is a schematic specification of swapping two qubits, which is implemented by the circuit on the left. Mathematically, the equality is described by {\tt Lemma Eq15} below.
	\begin{myverbatim}
	    \negskip
		Definition SWAP :=  B0 ⊗ B0 .+ B1 ⊗ B2 .+ B2 ⊗ B1 .+ B3 ⊗ B3.
		Definition XC :=  X ⊗ B3 .+ I_2 ⊗ B0.
		\negskip
		Lemma Eq15 : SWAP =  CX × XC × CX.
		\negskip
	\end{myverbatim} 
	In diagram (b), there is a controlled operation  performed on the second qubit, conditioned on the first qubit being set to zero. It is equivalent to a $CX$ gate enclosed by two Pauli-X gates on the first qubit. The  equality is specified by {\tt Lemma Eq16} below.\\[-2ex]
	\begin{myverbatim}	
	    \negskip
		Definition not_CX := B0 ⊗ X .+ B3 ⊗ I_2.
		\negskip
		Lemma Eq16 : not_CX = (X ⊗ I_2) × CX × (X ⊗ I_2).
	    \negskip
	\end{myverbatim} 
	In diagram (c), the controlled phase shift gate on the left is equivalent to a circuit for two qubits on the right. {\tt Lemma Eq17} gives a description of this equality.
	\begin{myverbatim}		
	    \negskip
		Definition CE (u: R) := B0 ⊗ I_2 .+ B3 ⊗ (Cexp u .* B0 .+ Cexp u .* B3).
		\negskip
		Lemma Eq17 : CE u = (B0 .+ Cexp u .* B3) ⊗ I_2.
		\negskip
	\end{myverbatim}
	In diagram (d), a controlled gate with two targets is equivalent to the concatenation of two $CX$ gates. 
	This is formalized by {\tt Lemma Eq18} below.
	\begin{myverbatim}		
	    \negskip
		Definition CXX := B0 ⊗ I_2 ⊗ I_2 .+ B3 ⊗ X ⊗ X.
		Definition CIX := B0 ⊗ I_2 ⊗ I_2 .+ B3 ⊗ I_2 ⊗ X.
		\negskip
		Lemma Eq18 : CXX = CIX × (CX ⊗ I_2).
		\negskip
	\end{myverbatim} \label{page:cix}
	The previous four lemmas can all be proved by using the strategy {\tt operate$\_$reduce} in conjunction with {\tt MatrixEquiv$\_$spec}.
	
\begin{figure}[t]
		\begin{minipage}[h]{0.45\linewidth}
			\[\Qcircuit @C=.8em @R=1.2em {
				& & & \\
				\lstick{x} & \qw & \gate{H} & \qw & \ctrl{2} & \qw & \qw & \\
				& & & & & & & \rstick{\beta_{xy}} \\
				\lstick{y} & \qw & \qw & \qw & \targ  & \qw & \qw & 
			}\]
			\caption{The  Bell states}\label{Bell}
		\end{minipage}\quad
		\begin{minipage}[h]{0.45\linewidth}	
			\[\begin{array}{rcl}
			|\beta_{00}\> & = & \frac{1}{\sqrt{2}} \sdot |0\> \otimes |0\> + \frac{1}{\sqrt{2}} \sdot |1\> \otimes |1\>\\
			|\beta_{01}\> & = & \frac{1}{\sqrt{2}} \sdot |0\> \otimes |1\> + \frac{1}{\sqrt{2}} \sdot |1\> \otimes |0\> \\
			|\beta_{10}\> & = & \frac{1}{\sqrt{2}} \sdot |0\> \otimes |0\> - \frac{1}{\sqrt{2}} \sdot |1\> \otimes |1\> \\
			|\beta_{11}\> & = & \frac{1}{\sqrt{2}} \sdot |0\> \otimes |1\> - \frac{1}{\sqrt{2}} \sdot |1\> \otimes |0\>
			\end{array}\]
		\end{minipage}
	\end{figure}
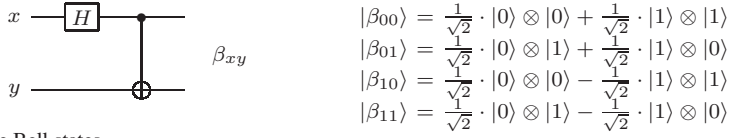

	In Section~\ref{sec:symbolic} we have formalized the preparation of the 3-qubit GHZ state (cf. Figure~\ref{GHZ}). Now let us  have a look at the Bell states. Depending on the input states, the circuit in Figure~\ref{Bell} gives four possible output states. The correctness of the circuit is validated by the four lemmas below, where the states are given in terms of density matrices and the circuit is described by a super-operator. It is easy to prove them by using our strategy {\tt super$\_$reduce}.
	
	\begin{myverbatim}
		\negskip
		Definition bl00 := /√2 .* (∣0,0⟩) .+ /√2 .* (∣1,1⟩).
		Definition bl01 := /√2 .* (∣0,1⟩) .+ /√2 .* (∣1,0⟩).
		Definition bl10 := /√2 .* (∣0,0⟩) .+ (-/√2) .* (∣1,1⟩).
		Definition bl11 := /√2 .* (∣0,1⟩) .+ (-/√2) .* (∣1,0⟩).
		\negskip
		Lemma pb00 : super (CX × (H ⊗ I_2)) (density ∣0,0⟩) = density bl00.
		Lemma pb01 : super (CX × (H ⊗ I_2)) (density ∣0,1⟩) = density bl01.
		Lemma pb10 : super (CX × (H ⊗ I_2)) (density ∣1,0⟩) = density bl10.
		Lemma pb11 : super (CX × (H ⊗ I_2)) (density ∣1,1⟩) = density bl11.
		\negskip
	\end{myverbatim}	
	
	\subsection{Observational equivalence}
	An alternative way of interpreting a quantum circuit without measurements is to consider it  as an operator that changes input quantum states to output states and abstracts away unobservable details. Therefore, we reuse the notion of matrix equivalence $\approx$ introduced in Section~\ref{sec:symbolic} and call it
	observational equivalence  for quantum circuits.
	%
	
	The rationale of using $\approx$ is that from an observational point of view, global phases are irrelevant to the observed properties of the physical system under consideration and can be ignored as far as quantum states are concerned. 
	
	The following lemma provides a functional view of observational equivalence. It is a counterpart of {\tt MatrixEquiv$\_$spec} given in Section~\ref{sec:mat}.
	Let $A, B$ be two operators, we have $A\approx B$ if and only if $A|\psi\> \approx B|\psi\>$ for any state $|\psi\>$. 
	\begin{myverbatim}
		\negskip
		Lemma ObsEquiv_operator: forall \{n\} (A B: Matrix n n),
   A ≈ B <-> (forall \ps: Matrix n 1, A × \ps ≈  B × \ps).
		\negskip
	\end{myverbatim}
	Furthermore, two states are  equal modulo a global phase, i.e. $|\psi\> \approx |\phi\>$ if and only if	their density matrices are exactly the same, i.e. $|\psi\>\<\psi| = |\phi\>\<\phi|$. We formally prove this property as it motivated us to introduce density matrices to represent quantum states in Section~\ref{sec:density}.
    \begin{myverbatim}
		\negskip
	Lemma ObsEquiv_state: forall \{n\} (\ps \ph: Matrix n 1),
   \ps ≈ \ph <-> \ps × (\ps†) = \ph × (\ph†) .
		\negskip
	\end{myverbatim}
	
	Although both matrix equivalence $=$ and  observational equivalence $\approx$  can be used for relating circuits, the former is strictly finer  than the latter in the sense that $A=B$ implies $A\approx B$ but not the other way around. Therefore, in the rest of the paper, we relate circuits by  $=$ whenever possible, as they are also related by $\approx$. 
	
	Moreover, it is not difficult to see that $=$ is a congruence relation. For example,  if $A, B$ are two quantum gates and $A=B$, then we can add a control qubit to form  controlled-$A$ and controlled-$B$ gates, which are still identified by $=$. However, the relation $\approx$ does not satisfy such kind of congruence property. To see this, note that $I\approx -I$. But the controlled-$I$ gate is quite different from controlled-($-I$): the latter transforms $\frac{1}{\sqrt{2}}(|00\> + |11\>)$ into $\frac{1}{\sqrt{2}}(|00\> - |11\>)$ whereas the former keeps it unchanged. In Section~\ref{sec:deu} we will see a concrete example of using $\approx$, where quantum states are identified by purposefully ignoring their global phases.

	\section{Problems from Coq's type system and our solution} \label{sec:prob}
	
	In principle, the Dirac notation is  fully symbolic, i.e. no matter how we formalize it, the relevant laws and their proofs should remain unchanged. However, it turns out that different design choices in the formalization do make a difference.
	
	In Coq, there are three kinds of equivalence between expressions: (1) syntactic equality, (2) $\beta\eta$-reduction, and (3) provable equivalence.
	Specifically, $a$ and $b$ are syntactically equal if they are the same Coq expression.
	If $a$ and $b$ are not the same Coq expression, it is still possible for them to be $\beta\eta$-reducible to the same term.\footnote{Coq is an extension of the lambda calculus. The $\beta\eta$-reduction here means the $\beta\eta$-reduction of the underlying lambda calculus.} In that case, $a$ and $b$ can be used interchangeably in Coq, i.e. if $P(a)$ is a well-typed proposition, then so is $P(b)$, and every proof 
    of $P(a)$  is also a proof of $P(b)$. Moreover, if $T(a)$ is a legal Coq type, then so is $T(b)$, and every element of $T(a)$ is also an element of $T(b)$.
	If $a$ and $b$ are not $\beta\eta$-equivalent, it is still possible for them to be provably equivalent. For example, the two expressions $(1+n)$ and $(n+1)$ are not $\beta\eta$-equivalent in Coq for a general variable $n$ but they are provably equal to each other. In other words, a proof of $P(1+n)$  is NOT necessarily a proof of $P(n+1)$ although we can always derive a proof of $P(n+1)$ from a proof of $P(1+n)$. Moreover, if $T(1+n)$ is a type, its element is not necessarily an element of type $T(n+1)$.
	
	\paragraph{Matrix definition.}
	When it comes to matrix definitions. The first problem is to decide whether $2^{n+1}\times 2^{n+1}$ matrices and $2^{1+n}\times 2^{1+n}$ matrices are $\beta\eta$-equivalent Coq types or not.
	Intuitively, since these two kinds of matrices are mathematically the same object, they should be used interchangeably. However, $(1+n)$ and $(n+1)$ are not $\beta\eta$-equivalent. Thus, we have to carefully define the Coq type of matrices so that those two kinds above are $\beta\eta$-equivalent types.
	We follow the approach used in QWIRE~\cite{PRZ17} and  define matrices (no matter how large they are) to be functions from two natural numbers (row and column numbers) to complex numbers.
	\begin{myverbatim}
		\negskip
		Definition Matrix (m n : N) := nat -> nat -> C.
		\negskip
	\end{myverbatim}
    But still, there are two problems that need to be solved.

	\paragraph{The elements outside the range of a matrix.}
	Intuitively, this definition above says that, given a pair of numbers $(k,l)$, if $k< m$ and $l< n$ then the entry of the matrix in the $k$-th row and $l$-th column is a complex number determined by the mapping. However, if $k\geq m$ or $l\geq n$ then the number determined by the mapping does not correspond to a valid entry in the matrix. In manual proofs we can simply ignore those elements outside the range of the matrix.  
    \begin{equation}\label{eq:M1}
    A_{m\times n} =
    \begin{bmatrix}
    a_{00} & \cdots & a_{0(n-1)} & 0  & \cdots  \\
    \vdots &   & \vdots & \vdots &   \\
    a_{(m-1)0} &  \cdots & a_{(m-1)(n-1)} & 0 & \cdots   \\
    0 & \cdots & 0 & 0  & \cdots   \\
    \vdots & & \vdots &   \vdots &   
    \end{bmatrix}
    \end{equation}
    
    \begin{equation}\label{eq:M2}
    B_{m\times n} =
    \begin{bmatrix}
    a_{00} & \cdots & a_{0(n-1)} & 1  & \cdots  \\
    \vdots &   & \vdots & \vdots &   \\
    a_{(m-1)0} &  \cdots & a_{(m-1)(n-1)} & 1 & \cdots   \\
    1 & \cdots & 1 & 1  & \cdots   \\
    \vdots & & \vdots &   \vdots &   
    \end{bmatrix}
    \end{equation}
    For example, in (\ref{eq:M1}) and (\ref{eq:M2}) we give two mappings in the form of two infinite-dimensional matrices $A_{m\times n}$ and $B_{m\times n}$, respectively.  Basically, $A_{m\times n}$ is the same as $B_{m\times n}$  except that all the elements with rows (resp. columns) greater than or equal to $m$ (resp. $n$) are $0$ in the former but they are $1$ in the latter.
    In computer-aided proofs, we could choose to only reason about {\it well-formed} matrices whose ``outside  elements'' are all zero like $A_{m\times n}$ above. Paykin et al.~\cite{PRZ17} heavily used this approach in their work. They would consider $A_{m\times n}$ well-formed but  $B_{m\times n}$ ill-formed.
	However, only reasoning about well-formed matrices imposes a heavy burden for formal proofs because the condition of well-formedness needs to be checked each time we manipulate matrices. In our development, we 
	choose a relaxed notion of matrix equivalence, which also appeared in~\cite{PRZ17},
	so that two matrices are deemed to be equivalent if they are equal component-wisely within the desired dimensions, and outside the  dimensions the corresponding elements can be different. For instance, this notion of equivalence allows us to identify $A_{m\times n}$ with $B_{m\times n}$. 
	With a slight abuse of notation, we still use the symbol $=$
	to denote the newly defined matrix equivalence\footnote{Nevertheless, we keep our Coq script in the repository at Github more rigid. There we use $\equiv$ to stand for the relaxed notion of matrix equivalence and reserve $=$ for the stronger notion of equivalence in the sense that $A=B$ means the two matrices $A$ and $B$ are equal component-wisely both within and outside their dimensions.}, and prove its elementary properties about scale product, matrix product, matrix addition, tensor product and conjugate transpose. 
	Reasoning about matrices modulo that equivalence turns out to be convenient in Coq. Specifically, the automation of the rewriting strategies mentioned above does not require side condition proofs about well-formedness.

	\paragraph{Coq type casting for rewriting.}
	\label{par:type}
	In math, $|0\>\otimes|0\>$ is a $4\times 1$ matrix and it is only verbose to say it is a $(2\cdot2) \, \times (1\cdot1)$ matrix. Even though $(2\cdot2) \, \times (1\cdot1)$ is convertible to $4\times 1$,
	these two typing claims are not syntactically identical in Coq but only $\beta\eta$-equivalent to each other. This difference is significant in rewriting.
	For example, the associativity of matrix multiplication is usually described as follows.
	$$A \times (B \times C) = (A \times B) \times C$$
	But more formally, the associativity means for any natural numbers $m$, $n$, $o$, $p$ and $m\times n$ matrix $A$, $n\times o$ matrix $B$ and $o\times p$ matrix $C$,
	$$A \underset{m,n,p}{\times} (B \underset{n,o,p}{\times} C) = (A \underset{m,n,o}{\times} B) \underset{m,o,p}{\times} C,$$
	where we use subscripts under $\times$ to indicate matrices' dimensions.
	These parameters do appear (implicitly) in Coq's formalization of matrix multiplication's associativity.
	Thus rewriting does not work in the following case
	$$A \underset{1,1,1}{\times} (B \underset{1\cdot 1,1,1\cdot 1}{\times} C), $$
	because rewriting uses an exact syntax match.
	This problem of type mismatch often occurs after we use the law {\law}13 for rewriting. We choose to build a customized rewrite tactic to overcome this problem.
	Using the example above, we want to rewrite via the associativity of multiplication. We first do a pattern matching for expressions of the form
	$$A \underset{m,n_1,p_1}{\times} (B \underset{n_2,o,p_2}{\times} C), $$
	no matter whether {\tt n1} and {\tt p1} coincide with {\tt n2} and {\tt p2}, respectively. We then use Coq's built-in unification to unify {\tt n1}, {\tt p1} with {\tt n2}, {\tt p2}. This unification must succeed or else the original expression of matrix computation is not well-formed. After the expression is changed to
	$$A \underset{m,n_1,p_1}{\times} (B \underset{n_1,o,p_1}{\times} C), $$
	we can use Coq's original rewrite tactic via the associativity of multiplication.
	
	We handle the above mentioned type problems silently and whoever uses our system to formalize his/her own proof will not even feel these problems.

	\section{Case studies}\label{sec:casestudy}
	To illustrate the power of our symbolic approach of reasoning about quantum circuits, we conduct a few case studies and compare the approach with the computational one in~\cite{PRZ17}.
	
	\subsection{Deutsch's algorithm}\label{sec:deu}
	Given a boolean function $f : \{0,1\} \rightarrow \{0,1\}$, Deutsch~\cite{Deu85} presented a quantum algorithm that can compute  $f(0)\oplus f(1)$ in a single evaluation of $f$. The algorithm can tell whether $f(0)$ equals $f(1)$ or not, without giving any information about the two values individually.
	The quantum circuit in Figure~\ref{Deu} gives an implementation of the algorithm. It makes use of a quantum oracle that maps any state $|x\>\otimes|y\>$ to the state $|x\>\otimes |y\oplus f(x)\>$, where $x,y\in\{0,1\}$.  More specifically, the unitary operator $U_f$ can be in one of the following four forms:
	\begin{itemize}
		\item if $f(0)=f(1)=0$, then $U_f=U_{f00}=I_2 \otimes I_2$;
		\item if $f(1)=f(1)=1$, then $U_f=U_{f11}=I_2\otimes X$;
		\item if $f(0)=0$ and $f(1)=1$, then $U_f=U_{f01}=CX$;
		\item if $f(0)=1$ and $f(1)=0$, then $U_f=U_{f10}=B_0 \otimes X + B_3 \otimes I_2$.
	\end{itemize}
	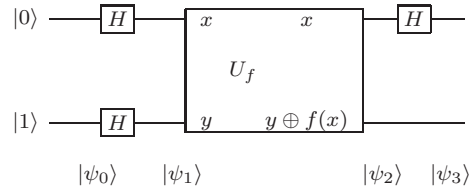
\begin{figure}
		
		\[\Qcircuit @C=0.8em @R=1.8em {
			\lstick{\ket{0}} & \qw & \qw  & \gate{H} & \qw & \qw  & \multigate{2}{\mathcal{
					\begin{array}{cccc}
					x & & & x \\
					& & &\\
					& & U_f &\\
					& & &\\
					y & & & y \oplus f(x) \\
					\end{array}%
			}} & \qw & \gate{H} & \qw & \qw \\
			& & &\\
			\lstick{\ket{1}} & \qw & \qw  & \gate{H} & \qw  & \qw & \ghost{\mathcal{
					\begin{array}{cccc}
					x & & & x \\
					& & &\\
					& & U_f &\\
					& & &\\
					y & & & y \oplus f(x) \\
					\end{array}%
			}} & \qw & \qw & \qw & \qw \\
			& \rstick{\ket{\psi_0}} & & & \rstick{\ket{\psi_1}} & & & & \lstick{\ket{\psi_2}} & & & \lstick{\ket{\psi_3}} }\]	
		
		\caption{Deutsch's algorithm}\label{Deu}
	\end{figure}
	
	We formalize Deutsch's algorithm in Coq and use our symbolic approach to prove its correctness. Let us suppose that  $|\psi_0\>= \ket{01}$ is the input state. There are three phases in this quantum circuit. The first phase  applies the Hadamard gate to each of the two qubits. So we define the initial state and express the state after the first phase as follows.
	\begin{myverbatim}
		\negskip
		Definition \ps0 := ∣0⟩ ⊗ ∣1⟩.
		Definition \ps1 := (H ⊗ H) × \ps0.
		\negskip
		Lemma step1 : \ps1 = ∣+⟩ ⊗ ∣-⟩.
		\negskip
	\end{myverbatim}
	Lemma {\tt step1} claims that the intermediate state after the first phase is $|+\> \otimes |-\>$. We can use the strategy {\tt operate$\_$reduce} designed in Section~\ref{sec:symbolic} to prove its correctness.
	
	The second phase applies  the unitary operator $U_f$ to $|\psi_1\>$. Since $U_f$ has four possible forms, we consider four cases. 
	\begin{myverbatim}
		\negskip
		Definition \ps20 := (I_2 ⊗ I_2) × \ps1.
		Definition \ps21 := (I_2 ⊗ X) × \ps1.
		Definition \ps22 := CX × \ps1.
		Definition \ps23 := (B0 ⊗ X .+ B3 ⊗ I_2) × \ps1.
		\negskip
		Lemma step20 : \ps20 = ∣+⟩ ⊗ ∣-⟩.
		Lemma step21 : \ps21 = -1 .* ∣+⟩ ⊗ ∣-⟩.
		Lemma step22 : \ps22 = ∣-⟩ ⊗ ∣-⟩.
		Lemma step23 : \ps23 = -1 .* ∣-⟩ ⊗ ∣-⟩.
		\negskip
	\end{myverbatim}
	Each of the above four lemmas corresponds to one case. They claim that the intermediate state $|\psi_{2}\>$ is $\pm1 \sdot |+\> \otimes |-\>$ after the second phase when $f(0) = f(1)$, and $\pm1 \sdot |-\> \otimes |-\>$ when $f(0) \neq f(1)$. We prove the four lemmas by rewriting $|\psi_1\>$ with Lemma {\tt step1} and using the strategy {\tt operate$\_$reduce} again. 
	
	The last phase applies the Hadamard gate to the first qubit of $|\psi_{2}\>$. So we still have four cases.
	\begin{myverbatim}
		\negskip
		Definition \ps30 := (H ⊗ I_2) × \ps20.
		...
		\negskip
		Lemma step30 : \ps30 = ∣0⟩ ⊗ ∣-⟩.
		Lemma step31 : \ps31 = -1 .* ∣0⟩ ⊗ ∣-⟩.
		Lemma step32 : \ps32 = ∣1⟩ ⊗ ∣-⟩.
		Lemma step33 : \ps33 = -1 .* ∣1⟩ ⊗ ∣-⟩.
		\negskip
	\end{myverbatim}
	Observe that the only difference between $|\psi_{30}\>$ and $|\psi_{31}\>$ lies in the global phase  $-1$, which can be ignored. Similarly for $|\psi_{32}\>$ and $|\psi_{33}\>$. Formally, we can prove the following lemmas.
	\begin{myverbatim}
	    \negskip
		Lemma step31' : \ps31 ≈ ∣0⟩ ⊗ ∣-⟩.
		Lemma step33' : \ps33 ≈ ∣1⟩ ⊗ ∣-⟩.
		\negskip
	\end{myverbatim}
	Therefore, after the last phase, we have $|\psi_{3}\>= |0\> \otimes |-\>$ when $f(0) = f(1)$, and $|\psi_{3}\>= |1\>\otimes |-\>$ when $f(0) \neq f(1)$. This is proved by using the intermediate results obtained in the first two phases and the strategy {\tt operate$\_$reduce}. 

	The above reasoning about the Deutsch's algorithm proceeds step by step and shows all the intermediate states in each phase. 
	Alternatively, one may be only interested in the output state of the circuit once an input state is fed. In other words, we would like to show a property like  $$|\psi_{3ij}\>= (H\otimes I_2)\times U_{fij}\times (H\otimes H) \times |\psi_0\>.$$ 
	Formally, we need to prove four equations depending on the forms of $U_f$.
	\begin{myverbatim}
		\negskip
		Lemma deutsch00 : 
   (H ⊗ I_2) × (I_2 ⊗ I_2) × (H ⊗ H) × (∣0⟩ ⊗ ∣1⟩) = ∣0⟩ ⊗ ∣-⟩ .
		Lemma deutsch01 : 
   (H ⊗ I_2) × (I_2 ⊗ X) × (H ⊗ H) × (∣0⟩ ⊗ ∣1⟩) = -1 .* ∣0⟩ ⊗ ∣-⟩ .
		Lemma deutsch10 : 
   (H ⊗ I_2) × CX × (H ⊗ H) × (∣0⟩ ⊗ ∣1⟩ = ∣1⟩ ⊗ ∣-⟩ .
		Lemma deutsch11 : 
   (H ⊗ I_2) × (B0 ⊗ X .+ B3 ⊗ I_2) × (H ⊗ H) × (∣0⟩ ⊗ ∣1⟩) = -1 .* ∣1⟩ ⊗ ∣-⟩.
		\negskip
	\end{myverbatim}
	The second and fourth equations can be written in a simpler form as follows.
	\begin{myverbatim}
		\negskip
		Lemma deutsch01' : 
   (H ⊗ I_2) × (I_2 ⊗ X) × (H ⊗ H) × (∣0⟩ ⊗ ∣1⟩) ≈ ∣0⟩ ⊗ ∣-⟩ .
		Lemma deutsch11' : 
   (H ⊗ I_2) × (B0 ⊗ X .+ B3 ⊗ I_2) × (H ⊗ H) × (∣0⟩ ⊗ ∣1⟩) ≈ ∣1⟩ ⊗ ∣-⟩ .
		\negskip
	\end{myverbatim}	
	Using our symbolic reasoning, these lemmas can be easily proved by {\tt rewrite ObsEquiv$\_$state} and {\tt operate$\_$reduce}. Thus, we know that the first qubit of the result state is $|0\>$ when $f(0) = f(1)$, and $|1\>$ otherwise. That is, $|\psi_{3ij}\>= |f(0) \oplus f(1)\> |-\>$ as expected.
	
	
	\subsection{Teleportation}\label{sec:tele}
	Quantum teleportation~\cite{BB93} is one of the most important protocols in quantum information theory. It teleports an unknown quantum state by only sending classical information, by making use of a maximally entangled state. In particular, the algorithm involves performing intermediate measurements on the quantum state and then applying different operations on the intermediate measurement results represented by  mixed states. 
	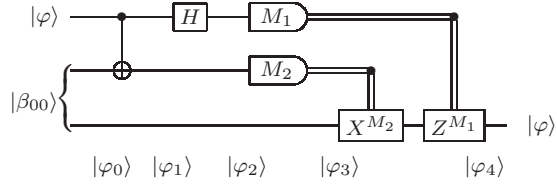
\begin{figure}[t]
		\[\Qcircuit @C=0.5em @R=1.2em {
			\lstick{\ket{\varphi}} & \qw & \qw & \qw & \ctrl{1} & \qw & \qw & \qw & \gate{H} & \qw & \qw & \qw & \measureD{M_1} & \cw & \cw & \cw & \cw & \control \cw  \cwx[2]  \\
			\lstick{} & \qw & \qw & \qw & \targ & \qw & \qw & \qw & \qw & \qw & \qw & \qw & \measureD{M_2} & \cw & \cw &  \control \cw  \cwx[1] \\
			\lstick{} & \qw & \qw & \qw &\qw & \qw &\qw & \qw & \qw & \qw & \qw & \qw & \qw & \qw & \qw & \gate{X^{M_2}} & \qw & \gate{Z^{M_1}} & \qw & \qw & \rstick{\ket{\varphi}} \inputgroupv{2}{3}{0.5em}{1.5em}{\ket{\beta_{00}}}\\
			& \rstick{\ket{\varphi_0}} & & & & \rstick{\ket{\varphi_1}} & & & & & & & \lstick{\ket{\varphi_2}} & & & \lstick{\ket{\varphi_3}} & & & & & \lstick{\ket{\varphi_4}}}\]
		\caption{Teleportation~\cite{NC11}}\label{Tele}
	\end{figure}
	
	Let the sender and the receiver  be $Alice$ and $Bob$, respectively. The quantum teleportation protocol goes as follows, as illustrated by the quantum circuit in Figure~\ref{Tele}.
	\begin{enumerate}
		\item $Alice$ and $Bob$ prepare an EPR state $|\beta_{00}\rangle_{q_2,q_3}$ together. Then they share the qubits, $Alice$ holding $q_2$ and $Bob$ holding $q_3$.
		\item To transmit the state $|\varphi\>$ of the quantum qubit $q_1$, $Alice$ applies a $CX$ operation on $q_1$ and $q_2$ followed by an $H$ operation on $q_1$.
		\item $Alice$ measures $q_1$ and $q_2$ and sends the outcome $x$ to $Bob$.
		\item When $Bob$ receives $x$, he applies appropriate Pauli gates 
		on his qubit $q_3$ to recover the original state $|\varphi\>$ of $q_1$.
	\end{enumerate}
								 				   
	We formalize the quantum teleportation protocol in Coq and use our symbolic approach to prove its correctness. Let $|\varphi\>=\alpha \ket{0}+\beta \ket{1}$ be any vector used as a part of the input state. The other part is $|\beta_{00}\>$, which needs extra preparation. For simplicity, we directly represent $|\beta_{00}\>$ with a combination of $|0\>$ and $|1\>$. So we define the input state $|\varphi_0\>$ as follows.
	\begin{myverbatim}
	    \negskip
		Variables (\al \be : C).
		Hypothesis Normalise: |\al|^2 + |\be|^2 = 1.
		Definition \var : Vector 2 := \al .* ∣0⟩ .+ \be .* ∣1⟩.
		\negskip
		Definition \var0 := \var ⊗ bl00.
		\negskip
	\end{myverbatim} 
	
	The input state goes through the quantum circuit that comprises four phases. We can easily define the quantum pure state $|\varphi_2\>$ after the second phase as follows, and prove Lemma {\tt tele1} by {\tt operate\_reduce}.
	\begin{myverbatim}
	    \negskip
		Definition \var2 := (H ⊗ I_2 ⊗ I_2) × (CX ⊗ I_2) × \var0.
		\negskip
		Lemma tele1 : \var2 = \al .* (∣+⟩ ⊗ bl00) .+ \be .* (∣-⟩ ⊗ bl01).
		\negskip
	\end{myverbatim} 
	
	In the third phase, due to the measurement with measurement operators $\{N_0,N_1\}$, where $N_0=B_0$ and $N_1=B_3$, there are four possible cases for the state $|\varphi_3\>$, and the probability for each case can be calculated as
\[\begin{array}{rcl}
\tr((N_i \otimes N_j\otimes I_2)^\dagger\times (N_i\otimes N_j\otimes I_2) \times |\varphi_2\>\<\varphi_2|) ~=~ 1/4.
\end{array}\]
	So we define $p_{ij}$ and $\rho_{3ij}$ as follows, where $i,j\in\{0,1\}$ are the measurement outcomes for the top two qubits. Then we prove Lemma {\tt tele2} about mixed states by {\tt super\_reduce} and some data processing. (We only consider projection operators as measurement operators, so we can replace $N ^\dag \times N$ with $N$).

	\begin{myverbatim}
	\negskip
	Definition p00 := trace ((B0 ⊗ B0 ⊗ I_2) × (density \var2)).
		Definition \rh300 := 1/p00 .* (super (B0 ⊗ B0 ⊗ I_2)(density \var2)).
		...
		Lemma tele2 : 
		  MeaMix 2 1 (MeaDen 2 0 (density \var2)) .=
		  [(1/4, 4 .* density (/√2 .* (∣00⟩ ⊗ (\al .* ∣0⟩ .+ \be .* ∣1⟩))));
		  [(1/4, 4 .* density (/√2 .* (∣01⟩ ⊗ (\al .* ∣1⟩ .+ \be .* ∣0⟩))));
		  [(1/4, 4 .* density (/√2 .* (∣10⟩ ⊗ (\al .* ∣0⟩ .+ -\be .* ∣1⟩))));
		  [(1/4, 4 .* density (/√2 .* (∣11⟩ ⊗ (\al .* ∣1⟩ .+ -\be .* ∣0⟩))))].
	\end{myverbatim} 	
	
	Finally, according to the different measurement results on the first two qubits, we apply corresponding  Pauli gates on the third qubit. So the quantum state after the fourth phase becomes $|\varphi_{4ij}\>$. We can formalize  it in Coq as follows.
	\[\begin{array}{rcl}
	|\varphi_{4ij}\> := (I_2 \otimes I_2 \otimes Z^i) \times (I_2 \otimes I_2 \otimes X^j) \times |\varphi_{3ij}\>
	\end{array}\]
	\begin{myverbatim}
	    \negskip
		Definition \rh400 := \rh30.
		Definition \rh401 := super (I_2 ⊗ I_2 ⊗ X) \rh310.
		Definition \rh410 := super (I_2 ⊗ I_2 ⊗ Z) \rh310.
		Definition \rh411 := super ((I_2 ⊗ I_2 ⊗ Z)×(I_2 ⊗ I_2 ⊗ X)) \rh311.
        \negskip
		Lemma tele3 : 
		  [(1/4, \rh400);[(1/4, \rh401); [(1/4, \rh410);[(1/4, \rh411)] .=
		  [(1/4, density (∣0,0⟩ ⊗ \ps); (1/4, density (∣0,1⟩ ⊗ \ps);
		  [(1/4, density (∣1,0⟩ ⊗ \ps); (1/4, density (∣1,1⟩ ⊗ \ps)].
	    \negskip
	\end{myverbatim} 	
	Using the simplification strategies discussed in Section~\ref{sec:sym}, it is easy to prove that $|\varphi_{4ij}\>$ can be simplified to be  $|i\>\otimes|j\>\otimes|\varphi\>$, and the probability of each case is $\frac{1}{4}$.

	In summary, we can use the following equality to express the whole protocol.
	\[\begin{array}{cl}
	|\varphi_{4ij}\>= (I_2 \otimes I_2 \otimes Z^i) \times (I_2 \otimes I_2 \otimes X^j) \times (N_i\otimes N_j\otimes I_2)\\
	\times
	(H\otimes I_2\otimes I_2)\times(CX \otimes I_2) \times (|\varphi\>\otimes |\beta_{00}\>).
	\end{array}
	\]
    It shows that the third qubit of the result state is always equal to $|\varphi\>$, the state to be teleported from Alice to Bob.

	\subsection{Simon's Algorithm}\label{sec:simon}
	The Simon's problem was raised in 1994~\cite{Simon97}. Although it is an artificial problem, it inspired Shor to discover a polynomial time algorithm to solve the integer factorization problem.
	
	Given a function $f : \{0,1\}^n \rightarrow \{0,1\}^n$, suppose there exists a string $s \in \{0,1\}^n$ such that the following property is satisfied
	\begin{equation}\label{eq:v}\begin{array}{cl}
			f(x) = f(y)  ~~\Leftrightarrow~~ x=y\ \mbox{ or }\ x \oplus y =s,
	\end{array} \end{equation}
	for all $x,y \in \{0,1\}^n$. 
	Here $\oplus$ is the bit-wise modulo 2 addition of two $n$ bit-strings. The goal of Simon's algorithm is to find the string $s$. The algorithm consists of iterating the quantum circuit
	and then performing some classical post-processing.
	
	\begin{enumerate}
		\item Set an initial state $|0\>^{\otimes n} \otimes |0\>^{\otimes n}$, and apply Hadamard gates to the first $n$ qubits respectively.
		\item Apply an oracle $U_f$ to all the $2n$ qubits, where  $U_f: |x\>|y\> \mapsto |x\>|f(x) \oplus y\>$. 
		\item Apply Hadamard gates to the first $n$ qubits again and then measure them. 
	\end{enumerate}
	
	When $s \neq 0^n$, the probability of obtaining each string $y\in\{0,1\}^n$ is
	$$ p_y =\left\{
	\begin{array}{ll}
	2^{-(n-1)} \qquad       & {\rm if} \quad s \sdot y = 0\\
	0           & {\rm if} \quad s \sdot y = 1.
	\end{array} \right. $$
	Therefore, it can be seen that the result string $y$ must satisfy $s \sdot y = 0$ and be evenly distributed. Repeating this process $n-1$ times, we will get $n-1$ strings $y_1,\cdots,y_{n-1}$ so that $y_i \cdot s=0$ for $1\leq i\leq n-1$. Thus we have  $n-1$ linear equations with $n$ unknowns ($n$ is the number of bits in s). The goal is to solve this system of equations to get $s$. We can get a unique non-zero solution $s$ if we are lucky and $y_1,...,y_{n-1}$ are linearly independent. Otherwise, we repeat the entire process and will find a linearly independent set with a high probability.
	
	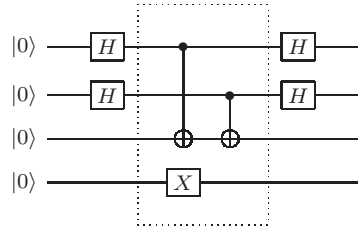
\begin{figure}
		\[\Qcircuit @C=1em @R=1em { 
			& & & & & & & & &\\
			\lstick{\ket{0}} & \qw	& \gate{H} & \qw & \ctrl{2} & \qw & \qw & \gate{H} & \qw &\qw \\
			\lstick{\ket{0}} & \qw	& \gate{H} & \qw & \qw & \ctrl{1} & \qw & \gate{H} & \qw &\qw \\
			\lstick{\ket{0}} & \qw	& \qw & \qw & \targ & \targ & \qw & \qw & \qw &\qw\\
			\lstick{\ket{0}} & \qw	& \qw & \qw & \gate{X} & \qw & \qw & \qw & \qw &\qw \\
			& & & & & & & & & \quad \gategroup{1}{4}{6}{7}{.7em}{.} \\
		}\]
		\caption{Simon's algorithm with $n$ = 2 and $s$ = 11~}\label{Simon11}
	\end{figure}
	
	As an example, we consider the Simon's algorithm with $n = 2$. The quantum circuit is displayed in Figure~\ref{Simon11}. We design the oracle as the gates in the dotted box $U_f = (I_2 \otimes CX \otimes I_2) \times (CIX \otimes X)$, where the gate $CIX$ is defined in page~\pageref{page:cix}. For this oracle, $s = 11$ satisfies property (\ref{eq:v}). The change of states can be seen as follows.
	\[\begin{array}{rl}
	|0000\> & \xrightarrow{H \otimes H \otimes I_2 \otimes I_2} |++\>|00\> \\
	&
	\xrightarrow{U_f} \frac{1}{2}[(|00\>+|11\>)|01\>+(|01\>+|10\>)|11\>]\\
	& 
	\xrightarrow{H \otimes H \otimes I_2 \otimes I_2} \frac{1}{2}[(|00\>+|11\>)|01\>+(|00\>-|11\>)|11\>]
	\end{array} \]
	We can establish the following lemma with our symbolic approach and use the strategy {\tt super$\_$reduce} to prove it.
	\begin{myverbatim}
		\negskip
		Lemma simon : super ((H ⊗ H ⊗ I_2 ⊗ I_2) × (I_2 ⊗ CX ⊗ I_2) × 
   (CIX ⊗ X) × (H ⊗ H ⊗ I_2 ⊗ I_2)) (density ∣0,0,0,0⟩)	= density 
   (/2 .* ∣0,0,0,1⟩ .+ /2 .* ∣1,1,0,1⟩ .+ /2 .* ∣0,0,1,1⟩ .+ -/2 .* ∣1,1,1,1⟩).
		\negskip
	\end{myverbatim}
	
	We analyze the cases where the last two qubits are in the state $|01\>$ or $|11\>$. The corresponding first two qubits are in $|00\>$ or $|11\>$, each occurs with equal probability. By property (\ref{eq:v}), it means that $x \oplus y = 00$ or $11$, so we obtain that $s$ = 11.

	\subsection{Grover's algorithm}	

	In this section we consider Grover's search algorithm
	. The algorithm starts from the initial state $|0\>^{\otimes n}$. It first uses $H^{\otimes n}$ (the $H$ gate applied to each of the $n$ qubits) to obtain a uniform superposition state, and then applies the Grover iteration repeatedly. An implementation of the Grover iteration has 
	four steps:
	\begin{enumerate}
		\item Apply the oracle $O$;
		\item Apply the Hadamard transform $H^{\otimes n}$;
		\item Perform a conditional phase shift on $|x\>$, if $|x\> \neq |0\>$;
		\item Apply the Hadamard transform $H^{\otimes n}$ again.
	\end{enumerate}
	Here the conditional phase-shift unitary operator in the third step is $2|0\>\<0|-I$. We can merge the last three steps as follows,
	\[\begin{array}{cl}
	H^{\otimes n} \times (2|0\>\<0|-I) \times H^{\otimes n} = 2|\phi\>\<\phi|-I
	\end{array} \]
	where $|\phi\> = \frac{1}{\sqrt{N}}\sum\limits_{x=0}^{N-1} |x\>$ with $N=2^n$. Therefore, the Grover iteration becomes $G=(2|\phi\>\<\phi|-I) \times O$.
	
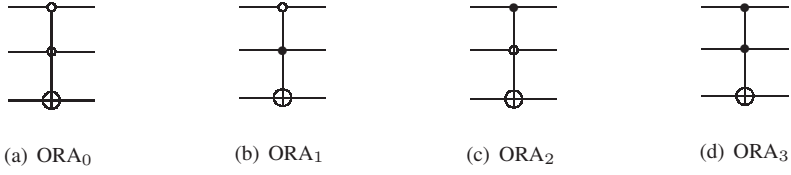
\begin{figure}
		\subfigure[ORA$_0$]{
			\begin{minipage}[t]{0.25\linewidth}
				\[\Qcircuit @C=.8em @R=1.7em {
					& \qw & \ctrlo{2} & \qw & \qw \\
					& \qw & \ctrlo{1} & \qw & \qw \\
					& \qw & \targ & \qw & \qw \\
					&  &  &  &  \\
				}\]			
			\end{minipage}%
		}%
		\subfigure[ORA$_1$]{
			\begin{minipage}[t]{0.25\linewidth}
				\[\Qcircuit @C=.8em @R=1.7em {
					& \qw & \ctrlo{2} & \qw & \qw \\
					& \qw & \ctrl{1} & \qw & \qw \\
					& \qw & \targ & \qw & \qw \\
					&  &  &  &  \\
				}\]	
			\end{minipage}%
		}%
		\subfigure[ORA$_2$]{
			\begin{minipage}[t]{0.25\linewidth}
				\[\Qcircuit @C=.8em @R=1.7em {
					& \qw & \ctrl{2} & \qw & \qw \\
					& \qw & \ctrlo{1} & \qw & \qw \\
					& \qw & \targ & \qw & \qw \\
					&  &  &  &  \\
				}\]	
			\end{minipage}%
		}%
		\subfigure[ORA$_3$]{
			\begin{minipage}[t]{0.25\linewidth}
				\[\Qcircuit @C=.8em @R=1.7em {
					& \qw & \ctrl{2} & \qw & \qw \\
					& \qw & \ctrl{1} & \qw & \qw \\
					& \qw & \targ & \qw & \qw \\
					&  &  &  &  \\
				}\]	
			\end{minipage}
		}%
		\caption{The quantum circuit of different Oracle}\label{G3}
	\end{figure}

	As a concrete example, we consider the Grover's algorithm with two qubits. The size of the search space of this algorithm is four. So we need to consider four search cases with $x^* = 0,1,2,3$. The oracle must satisfy that if $x=x^*$, then $f(x^*)=1$, otherwise $f(x)=0$. So in accordance with $x^* = 0,1,2,3$, we design four oracles $ORA_0, ..., ORA_3$, which are implemented by the four circuits in Figure~\ref{G3}.
	
	\begin{myverbatim}
		\negskip
		Definition ORA0 := B0 ⊗ (B0 ⊗ X .+ B3 ⊗ I_2) .+ B3 ⊗ I_2 ⊗ I_2.
		Definition ORA1 := B0 ⊗ CX .+ B3 ⊗ I_2 ⊗ I_2.
		Definition ORA2 := B0 ⊗ I_2 ⊗ I_2 .+ B3 ⊗ (B0 ⊗ X .+ B3 ⊗ I_2).
		Definition ORA3 := B0 ⊗ I_2 ⊗ I_2 .+ B3 ⊗ CX.
		\negskip
	\end{myverbatim}
	
	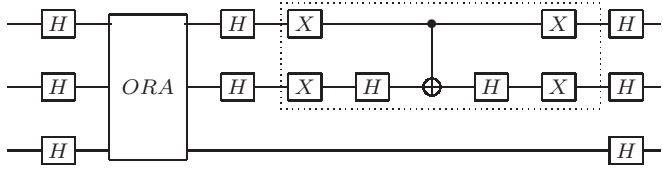
\begin{figure}
		\[\Qcircuit @C=.8em @R=1.7em {
			& \qw & \gate{H} & \qw & \multigate{2}{ORA} & \qw & \gate{H} & \qw & \gate{X} & \qw & \qw & \qw & \ctrl{1} & \qw & \qw & \qw & \gate{X} & \qw & \gate{H} & \qw\\
			& \qw & \gate{H} & \qw & \ghost{\mathcal{ORA}} & \qw & \gate{H} & \qw & \gate{X} & \qw & \gate{H} & \qw & \targ & \qw & \gate{H} & \qw & \gate{X} & \qw & \gate{H} & \qw\\
			& \qw & \gate{H} & \qw & \ghost{\mathcal{ORA}} & \qw & \qw & \qw & \qw & \qw & \qw & \qw & \qw & \qw & \qw & \qw & \qw & \qw & \gate{H} & \qw \gategroup{1}{9}{2}{18}{.7em}{.}
		}\]
		\caption{the Grover's algorithm with two qubits}\label{G4}
	\end{figure}
	
	The whole algorithm is illustrated by the circuit in Figure~\ref{G4}.
	The gates in the dotted box perform the conditional phase shift operation
	$2|0\>\<0|-I$. We then merge the front and back $H \otimes H$ gates to it and get the operation $CPS$ as follows.  
	\begin{myverbatim}
		\negskip
		Definition MI := (B0 .+ B1 .+ B2 .+ B3) ⊗ (B0 .+ B1 .+ B2 .+ B3).
		Definition CPS := (((/2 .* MI) .+ (-1) .* (I_2 ⊗ I_2)) ⊗ I_2).
		\negskip
	\end{myverbatim}
	So we have the Grover iteration $G=(2|\phi\>\<\phi|-I) \times O = CPS \times ORA_i$. Let the initial state be $|0\> \otimes |0\> \otimes |1\>$. After the Hadamard transform $H^{\otimes3}$, we only perform Grover iteration once to get the search solution. In summary, we formalize the Grover's algorithm with two qubits in the vector form as follows, and use {\tt operate$\_$reduce} to prove them. The reasoning using density matrices can also be done.
	\begin{myverbatim}
		\negskip
		Lemma Gro0:
   (I_2 ⊗ I_2 ⊗ H) × CPS × ORA0 × (H ⊗ H ⊗ H) × ∣0,0,1⟩ = ∣0,0,1⟩.
		Lemma Gro1:
   (I_2 ⊗ I_2 ⊗ H) × CPS × ORA1 × (H ⊗ H ⊗ H) × ∣0,0,1⟩ = ∣0,1,1⟩.
		Lemma Gro2:
   (I_2 ⊗ I_2 ⊗ H) × CPS × ORA2 × (H ⊗ H ⊗ H) × ∣0,0,1⟩ = ∣1,0,1⟩.
		Lemma Gro3:
   (I_2 ⊗ I_2 ⊗ H) × CPS × ORA3 × (H ⊗ H ⊗ H) × ∣0,0,1⟩ = ∣1,1,1⟩.
		\negskip
	\end{myverbatim}
	
	For all the case studies in the current and previous subsections, we employ the same schema: first represent the quantum circuits and input states symbolically, and then apply the strategies developed in Section~\ref{sec:sym} to show that the output states are equal to  our desired states. When the states are in the vector form, we use the strategy {\tt operate$\_$reduce}. When the states are in the density matrix form, we make use of the strategy {\tt super$\_$reduce}, which  decomposes density matrices into multiplications of vectors and then essentially resort to  {\tt operate$\_$reduce}, as discussed in Section~\ref{sec:sym}. In addition, if it is necessary to identify states up to an ignorance of global phases, we have the extra task of rewriting terms with  {\tt ObsEquiv$\_$state}.
	
	\subsection{Deutsch-Jozsa algorithm family}\label{sec:DJ}
	Representing a family of algorithms with unbounded qubit size, the Deutsch-Jozsa algorithm is a generalization application of the Deutsch algorithm on $n$ qubits. Given a boolean function $f : \{0,1\}^n \rightarrow \{0,1\}$ as a quantum oracle, the algorithm judges whether $f$ is a constant or a balanced function. Here $f(x)$ is balanced if it is equal to 1 for half of the input $x$, and 0 for the other half. 
	As shown in Figure~\ref{DJ}, the overall steps of the Deutsch-Jozas algorithm are as follows:
	\begin{figure}
	\begin{minipage}[h]{0.6\linewidth}
		\centering
		\[\Qcircuit @C=0.8em @R=1.8em {
			\lstick{\ket{0}} & \qw & {/^n}\qw & \qw & \gate{H^{\otimes}} & \qw & \qw  & \multigate{2}{\mathcal{
					\begin{array}{cccc}
					x & & & x \\
					& & &\\
					& & U_f &\\
					& & &\\
					y & & & y \oplus f(x) \\
					\end{array}%
			}} & \qw & \gate{H^{\otimes}} & \qw & \qw \\
			& & &\\
			\lstick{\ket{1}} & \qw & \qw & \qw  & \gate{H} & \qw  & \qw & \ghost{\mathcal{
					\begin{array}{cccc}
					x & & & x \\
					& & &\\
					& & U_f &\\
					& & &\\
					y & & & y \oplus f(x) \\
					\end{array}%
			}} & \qw & \qw & \qw & \qw \\
			& \rstick{\ket{\psi_0}} & & & & \rstick{\ket{\psi_1}} & & & & \lstick{\ket{\psi_2}} & & & \lstick{\ket{\psi_3}} }\]	
		\caption{Deutsch-Jozsa algorithm}\label{DJ}
	\end{minipage}\quad
	\begin{minipage}[h]{0.3\linewidth}	
		\centering
			\[\Qcircuit @C=1.0em @R=.8em {
	& \qw & \ctrl{1} & \qw & \qw & \qw & \qw & \qw & \qw & \qw & \qw & \ctrl{1} & \qw & \\
	& \qw & \targ  & \ctrl{1} & \qw & \qw & \qw & \qw & \qw & \qw & \ctrl{1} & \targ & \qw & \\
	& \qw & \qw & \targ & \qw & \qw & \qw & \qw & \qw & \qw & \targ& \qw & \qw & \\
	& & & & \cdots & & & & \cdots & & & & \\
	& \qw & \qw & \qw & \qw & \ctrl{1} & \qw & \ctrl{1} & \qw & \qw & \qw & \qw & \qw & \\
	& \qw & \qw & \qw & \qw & \targ & \ctrl{1} & \targ & \qw & \qw & \qw & \qw & \qw & \\
	& \qw & \qw & \qw & \qw & \qw & \targ & \qw & \qw & \qw & \qw & \qw & \qw &  \\
 			}\]
 		\caption{The circuit of $U_f$}\label{$U_f$}
	\end{minipage}
	\end{figure}

	\begin{enumerate}
	\item Apply Hadamard gates to 
	all the $n+1$ qubits respectively;
	\item Apply the oracle $U_f$ to the $n+1$ qubits, where  $U_f: |x\>|y\> \mapsto |x\>|y \oplus f(x)\>$; 
	\item Apply Hadamard gates to the first $n$ qubits again and then measure them.
	\end{enumerate}	
	
	Suppose the input state is $|\psi_0\> =|0\>^{\otimes n}|1\>$. After each of the above three steps, we get three states $|\psi_1\>$, $|\psi_2\>$ and $|\psi_3\>$ respectively.
		\[\begin{array}{rcl}
	|\psi_1\> & := & \sum\limits_{x=0}^{2^{n}-1} \frac{|x\>}{\sqrt{2^n}} |-\>\\
	|\psi_2\> & := &  \sum\limits_{x=0}^{2^{n}-1} \frac{(-1)^{f(x)}|x\>}{\sqrt{2^n}} |-\>\\
	|\psi_3\> & := & \sum\limits_{x=0}^{2^{n}-1} \sum\limits_{z=0}^{2^{n}-1} \frac{(-1)^{f(x)+x \sdot z}|z\>}{2^n} |-\>
	\end{array}\]
	From $H^{\otimes n} |0\>^{\otimes n} = \sum_{x} |x\>/\sqrt{2^n}$
	, we easily get $|\psi_1\>$. Then due to the fact that $U_f |-\> = (-1)^{f(x)}|-\>$, we have $|\psi_2\>$.	
	Finally, for a single qubit we know that $H|x\> = \sum_{z} {(-1)^{xz}|z\>}/\sqrt{2}$
	, we obtain $|\psi_3\>$, where $x \sdot z$ is the bitwise inner product of $x$ and $z$. After measuring the first $n$ qubits, we analyse the probability of the event that the final result is $|0\>^{\otimes n}$. When $|z\>$ is $|0\>$, the coefficient of $|\psi_3\>$ is $\sum_x (-1)^{f(x)+x \sdot z}/{2^n} |_{z=0}$
	, which is equal to $\sum_x (-1)^{f(x)}/{2^n}$
	. Thus, when $f$ is a constant function, the measurement result must be $|0\>^{\otimes n}$. Otherwise, when $f$ is a balanced one, the result would not be $|0\>^{\otimes n}$. In summary, based on the measurement of the first $n$ qubits, we can determine whether $f$ is a constant or a balanced function and only query the oracle once.
	
	In particular, consider the circuit of $U_f$ shown in Figure~\ref{$U_f$}, we specify $U_f$ in Coq as follows. 
	\begin{myverbatim}
		\negskip
		Fixpoint Uf (n:nat): Matrix (2*2^(N.of_nat n)) (2*2^(N.of_nat n)):=
   match n with
   | O => I 2
   | S n' => (CX ⊗ I (2^(N.of_nat n'))) × (I 2 ⊗ (Uf n')) 
           × (CX ⊗ I (2^(N.of_nat n')))
   end.
		\negskip
	\end{myverbatim}
	For convenience, we predefine an auxiliary function $kron\_n \ n \ A$, which stands for  $A^{\otimes n}$. When $n$ is 1, it degenerates into  $I_1$, i.e. 1. 
	\begin{myverbatim}
		\negskip
		Fixpoint kron_n (n:nat) {m1 m2} (A : Matrix m1 m2) 
   : Matrix (m1^(N.of_nat n)) (m2^(N.of_nat n)) :=
   match n with
   | 0    => I 1
   | S n' => kron A (kron_n n' A)
   end.
		\negskip
	\end{myverbatim}
	According to the above three steps, we use the following three lemmas to prove the correctness of the Deutsch-Jozsa algorithm.
	\begin{myverbatim}
	\negskip
Lemma DJ_0 :
   ((kron_n n H) ⊗ H) × ((kron_n n ∣0⟩) ⊗ ∣1⟩) = (kron_n n ∣+⟩) ⊗ ∣-⟩.
	\negskip
Lemma DJ_1 :
   (n > 0)%nat ->
   (Uf n) × ((kron_n n ∣+⟩) ⊗ ∣-⟩) = (kron_n n ∣+⟩) ⊗ ∣-⟩.
	\negskip
Lemma DJ_2 :
   ((kron_n n H) ⊗ H) × ((kron_n n ∣+⟩) ⊗ ∣-⟩) = (kron_n n ∣0⟩) ⊗ ∣1⟩.
	\negskip
	\end{myverbatim} 

	By using the proof assistant Coq, we can formalise the proof about $n$ qubits with induction and rewriting strategies. Since the number $n$ is unbounded, the correctness of the algorithm is difficult to be proved without proof assistant platforms. As an example, we consider the second lemma. We first focus on the case of $n=1$,
	\[\begin{array}{cl}
		CX \times (I_2 \otimes I_2) \times CX \times (|+\> \otimes |-\>) = |+\> \otimes |-\>.
	\end{array} \]	
	This case can be easily proved by {\tt operate$\_$reduce}. According to the inductive hypothesis, we assume that the algorithm is correct for the case of $n = k$. Then we are going to prove the case of $n=k+1$ with it and some laws in Table~\ref{t:core}. The proof steps are as follows.
	\[\begin{array}{cl}
	& U_{f}^{k+1} \times (|+\>^{\otimes (k+1)} \otimes |-\>)  \\
 	= & (CX \otimes I_{2^{k}}) \times (I_2 \otimes U_{f}^{k}) \times (CX \otimes I_{2^{k}}) \times (|+\> \otimes |+\> \otimes |+\>^{\otimes (k-1)} \otimes |-\>) \\
	= & (CX \otimes I_{2^{k}}) \times (I_2 \otimes U_{f}^{k}) \times ((CX \otimes I_{2^{k}}) \times ((|+\> \otimes |+\>) \otimes (|+\>^{\otimes (k-1)} \otimes |-\>))) \\
	= & (CX \otimes I_{2^{k}}) \times (I_2 \otimes U_{f}^{k}) \times ((CX \times (|+\> \otimes |+\>)) \otimes (I_{2^{k}} \times (|+\>^{\otimes (k-1)} \otimes |-\>))) \\
	= & (CX \otimes I_{2^{k}}) \times ((I_2 \otimes U_{f}^{k}) \times ((|+\> \otimes |+\>) \otimes (|+\>^{\otimes (k-1)} \otimes |-\>))) \\
	= & (CX \otimes I_{2^{k}}) \times ((I_2  \times |+\>) \otimes (U_{f}^{k} \times (|+\>^{\otimes k} \otimes |-\>))) \\
	= & (CX \otimes I_{2^{k}}) \times  (|+\> \otimes (|+\>^{\otimes k} \otimes |-\>)) \\
	= & (CX \otimes I_{2^{k}}) \times  ((|+\> \otimes |+\>) \otimes (|+\>^{\otimes k-1} \otimes |-\>)) \\
	= & (CX \times (|+\> \otimes |+\>)) \otimes (I_{2^{k}} \times (|+\>^{\otimes k-1} \otimes |-\>)) \\
	= & |+\>^{\otimes (k+1)} \otimes |1\> \\
	\end{array} \]
	
	After applying the Hadamard gates to the first $n$ qubits, we can get that the first $n$ qubits of the result is $|0\>^{\otimes n}$, So we know that the $f$ described as $U_f$ is a constant function.

	\subsection{Preparation of an entangled state}
\begin{figure}
		\begin{minipage}[h]{0.45\linewidth}
			\[\Qcircuit @C=1.0em @R=1.0em {
				& \gate{H} & \ctrl{1} & \ctrl{2} & \qw & \qw & \qw & \qw & \qw  & \qw & \qw & \\
				& \qw & \targ  & \ctrl{1} & \ctrl{2} & \qw & \qw & \qw & \qw  & \qw & \qw & \\
				& \qw & \qw & \targ & \ctrl{1} & \qw & \qw & \qw & \qw  & \qw & \qw & \\
				& \qw & \qw & \qw & \targ & \qw & \qw & \qw  & \qw & \qw & \qw & \\
				& & & & & & \cdots & & & & \\
				& \qw & \qw & \qw & \qw &\qw & \qw & \qw  & \ctrl{2} & \qw & \qw & \\
				& \qw & \qw & \qw & \qw &\qw & \qw & \qw  & \ctrl{1} & \ctrl{2} & \qw & \\
				& \qw & \qw & \qw & \qw & \qw & \qw & \qw  & \targ & \ctrl{1} & \qw & \\
				& \qw & \qw & \qw & \qw & \qw & \qw & \qw  & \qw & \targ & \qw & \\ 
			}\]
			\caption{Preparing an entangled state}\label{Example}
		\end{minipage}\quad
		\begin{minipage}[h]{0.45\linewidth}	
			\[\begin{array}{rcl}
			|000..00\>& \longrightarrow & [\frac{1}{\sqrt{2}}(|0\>+|1\>)] \otimes|00..00\>\\
			& \equiv & \frac{1}{\sqrt{2}}(|000..00\>+|100..00\>)\\
			& \longrightarrow & \frac{1}{\sqrt{2}}(|000..00\>+|110..00\>)\\
			& \longrightarrow & \frac{1}{\sqrt{2}}(|000..00\>+|111..00\>)\\
			&   &  \cdots \\
			& \longrightarrow & \frac{1}{\sqrt{2}}(|000..00\>+|111..10\>)\\
			& \longrightarrow & \frac{1}{\sqrt{2}}(|000..00\>+|111..11\>)\\
			\end{array}\]
		\end{minipage}
	\end{figure}
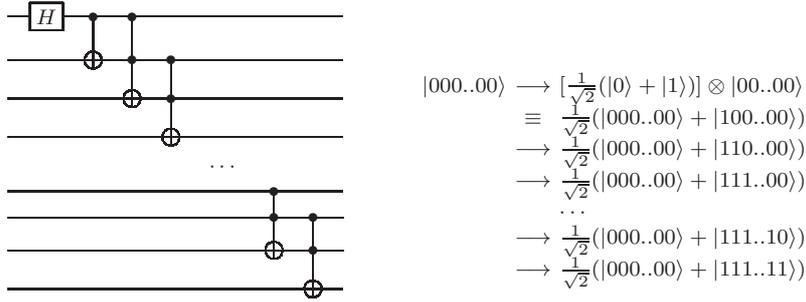

	So far we have seen some  simple examples with only a few qubits. 
	Now let us consider a bigger example with a dozen qubits. The quantum circuit  in Figure~\ref{Example} can be used to create a maximally entangled state.
	
	We start with the initial state $|000..00\>$ with $12$ qubits. First, we apply the H gate on the first qubit $q_1$ to change the global state into $[\frac{1}{\sqrt{2}}(|0\>+|1\>)] \otimes|00..00\>$ with  the first qubit in a superposition. It is equivalent to $\frac{1}{\sqrt{2}}(|000..00\>+|100..00\>)$. Then we apply the CX gate on the first and second qubits $q_1$ and $q_2$, which leads to the state $\frac{1}{\sqrt{2}}(|000..00\>+|110..00\>)$. Then we apply the Toffoli gate on the first three qubits to yield $\frac{1}{\sqrt{2}}(|000..00\>+|111..00\>)$. In a similar way, we  apply the Toffoli gate on the  qubits $q_{i}$, $q_{i+1}$ and $q_{i+2}$ for $i \in {2,3,..., n-2}$ in turn. Eventually, we obtain a maximally entangled state $\frac{1}{\sqrt{2}}(|000..00\>+|111..11\>)$.
	
The correctness of the quantum circuit is stated by the following lemma.
\begin{myverbatim}
	\negskip
	Lemma Entangled_state_12 :
   (I_2 ⊗ I_2 ⊗ I_2 ⊗ I_2 ⊗ I_2 ⊗ I_2 ⊗ I_2 ⊗ I_2 ⊗ I_2 ⊗ TOF )
   × (I_2 ⊗ I_2 ⊗ I_2 ⊗ I_2 ⊗ I_2 ⊗ I_2 ⊗ I_2 ⊗ I_2 ⊗ TOF ⊗ I_2)
   × (I_2 ⊗ I_2 ⊗ I_2 ⊗ I_2 ⊗ I_2 ⊗ I_2 ⊗ I_2 ⊗ TOF ⊗ I_2 ⊗ I_2)
   × (I_2 ⊗ I_2 ⊗ I_2 ⊗ I_2 ⊗ I_2 ⊗ I_2 ⊗ TOF ⊗ I_2 ⊗ I_2 ⊗ I_2)
   × (I_2 ⊗ I_2 ⊗ I_2 ⊗ I_2 ⊗ I_2 ⊗ TOF ⊗ I_2 ⊗ I_2 ⊗ I_2 ⊗ I_2)
   × (I_2 ⊗ I_2 ⊗ I_2 ⊗ I_2 ⊗ TOF ⊗ I_2 ⊗ I_2 ⊗ I_2 ⊗ I_2 ⊗ I_2)
   × (I_2 ⊗ I_2 ⊗ I_2 ⊗ TOF ⊗ I_2 ⊗ I_2 ⊗ I_2 ⊗ I_2 ⊗ I_2 ⊗ I_2)
   × (I_2 ⊗ I_2 ⊗ TOF ⊗ I_2 ⊗ I_2 ⊗ I_2 ⊗ I_2 ⊗ I_2 ⊗ I_2 ⊗ I_2)
   × (I_2 ⊗ TOF ⊗ I_2 ⊗ I_2 ⊗ I_2 ⊗ I_2 ⊗ I_2 ⊗ I_2 ⊗ I_2 ⊗ I_2)
   × (TOF ⊗ I_2 ⊗ I_2 ⊗ I_2 ⊗ I_2 ⊗ I_2 ⊗ I_2 ⊗ I_2 ⊗ I_2 ⊗ I_2)
   × (CX ⊗ I_2 ⊗ I_2 ⊗ I_2 ⊗ I_2 ⊗ I_2 ⊗ I_2 ⊗ I_2 ⊗ I_2 ⊗ I_2 ⊗ I_2)
   × (H ⊗ I_2 ⊗ I_2 ⊗ I_2 ⊗ I_2 ⊗ I_2 ⊗ I_2 ⊗ I_2 ⊗ I_2 ⊗ I_2 ⊗ I_2 ⊗ I_2)
   × ∣0,0,0,0,0,0,0,0,0,0,0,0⟩
   =≡ /√2 .* ∣0,0,0,0,0,0,0,0,0,0,0,0⟩ .+ /√2 .* ∣1,1,1,1,1,1,1,1,1,1,1,1⟩.
\end{myverbatim}
	By just using the strategy {\tt operate\_reduce}, we can prove the above property in half an hour. Note that this example cannot be handled by the computational approach --- going beyond $6$ qubits is very difficult for that approach.

	\subsection{Experiments}\label{sec:exp}
	\begin{table}[t]
		\begin{tabular}{|p{1.7cm}|p{1cm}|p{1cm}|p{1.35cm}|p{1.6cm}|p{1cm}|p{1cm}|}
			\hline  
			& Deutsch & Simon &Teleportation & Secret Sharing & QFT & Grover  \\  
			\hline  
			Symbolic  & 3656 & 53795 & 39715 & 68919 & 25096 & 146834 \\ 
			\hline
			Computational  & 25190 & 180724 & 46450 & 170490 & 68730 & 934570 \\   
			\hline  
		\end{tabular}
		\caption{Comparison of two approaches with verification time in milliseconds}\label{t:result}
	\end{table}
	
	We have conducted experiments on Deutsch's algorithm, Simon's algorithm,  quantum teleportation, quantum secret sharing protocol, quantum Fourier transform (QFT) with three qubits, and Grover's search algorithm with two qubits. In Table~\ref{t:result}, we record the execution time of those examples in milliseconds in CoqIDE 8.10.0 running on a PC with Intel Core  i5-7200 CPU and 8 GB RAM. As we can see in the table, our symbolic approach always outperforms  the computational one in~\cite{PRZ17}. 
	
	The computational approach is slow because of the explicit representation of matrices and inefficient tactics for evaluating matrix multiplications. 
	Let us consider a simple example.
	In the computational approach, the Hadamard gate $H$ is defined by {\tt ha} below:
	\begin{myverbatim}
		\negskip
		Definition ha : Matrix 2 2 := 
   fun x y => match x, y with
   | 0, 0 => (1 / √2)
   | 0, 1 => (1 / √2)
   | 1, 0 => (1 / √2)
   | 1, 1 => -(1 / √2)
   | _, _ => 0
   end.
		\negskip
	\end{myverbatim}
	Since $H$ is unitary, we have $HH = I$ and the following property becomes straightforward. 
	\begin{myverbatim}
		\negskip
		Lemma H3_ket0: (ha ⊗ ha ⊗ ha) × (ha ⊗ ha ⊗ ha) × (∣0,0,0⟩) = (∣0,0,0⟩).
		\negskip
	\end{myverbatim}	
	However, to prove the above lemma with the computational approach is far from being trivial.	
	To see this, we literally go through a few steps.
	Firstly, we apply the associativity of matrix multiplication  on the left hand side of the equation so to rewrite it into
	\[\begin{array}{rl}
	& (H \otimes H \otimes H) \times ((H \otimes H \otimes H) \times (|0\> \otimes |0\> \otimes |0\>)) .
	\end{array} \]	
	Secondly, each explicitly represented matrix is converted into a two-dimensional list and matrix multiplications are calculated in order. 
	Finally, we need to show that each of the eight elements in the vector on the left is equal to the corresponding element on the right.
	Let $A_0 = (H \otimes H \otimes H) \times (|0\> \otimes |0\> \otimes |0\>)$ and $A_1 = (H \otimes H \otimes H)  \times A_0$.
	With the computational approach, obvious simplifications such as multiplication and addition with $0$ and $1$ are carried out for 
	the elements in $A_0$ and $A_1$, and no more complicated simplification is effectively handled. So $A_0$ is a two-dimensional list with each element in the form $\frac{1}{\sqrt{2}} \times \frac{1}{\sqrt{2}} \times \frac{1}{\sqrt{2}}$ and $A_1$ is a two-dimensional list whose first element is 
	\[\begin{array}{l}
	(\frac{1}{\sqrt{2}} \times \frac{1}{\sqrt{2}} \times \frac{1}{\sqrt{2}} \times (\frac{1}{\sqrt{2}} \times \frac{1}{\sqrt{2}} \times \frac{1}{\sqrt{2}}))  \\ + (\frac{1}{\sqrt{2}} \times \frac{1}{\sqrt{2}} \times \frac{1}{\sqrt{2}} \times (\frac{1}{\sqrt{2}} \times \frac{1}{\sqrt{2}} \times \frac{1}{\sqrt{2}}))  \\
	+ \ ...  \\
	+ (\frac{1}{\sqrt{2}} \times \frac{1}{\sqrt{2}} \times \frac{1}{\sqrt{2}} \times (\frac{1}{\sqrt{2}} \times \frac{1}{\sqrt{2}} \times \frac{1}{\sqrt{2}})),\end{array}\] which is a summation of eight identical summands with $\frac{1}{\sqrt{2}}$ multiplied with itself six times; other elements are in  similar forms.  From this simple example, we can already see that the explicit matrix representation and ineffective simplification in matrix multiplication make the intermediate expressions very cumbersome. 
	
	On the contrary, in the symbolic approach we have 
	\[\begin{array}{rcl}
	A_1 & = & (H \otimes H \otimes H) \times (H \otimes H \otimes H) \times (|0\> \otimes |0\> \otimes |0\>)\\
	& = & (H \otimes (H \otimes H)) \times ((H \otimes (H \otimes H)) \times (|0\> \otimes (|0\> \otimes |0\>)))\\
	& = & (H \otimes (H \otimes H)) \times ((H \times |0\>) \otimes ((H \times |0\>) \otimes (H \times |0\>))) \\
	& = & (H \otimes (H \otimes H)) \times (|+\> \otimes (|+\> \otimes |+\>)) \\
	& = & (H \times |+\>) \otimes ((H \times |+\>) \otimes (H \times |+\>)) \\
	& = & |0\> \otimes (|0\> \otimes |0\>).\end{array}\] 
	Notice that here we have kept the structure of tensor products rather than to eliminate them. In fact, we lazily evaluate tensor products because they are expensive to calculate and  preserving more higher-level structures opens more opportunities for rewriting.
	The symbolic reasoning not only renders the intermediate expressions more readable, but also greatly reduces the time cost of arithmetic calculations.
	
	In general, in the computational approach a multiplication of two  $N \times  N$ matrices of $O(k)$-length expressions results in a matrix of $O(Nk)$-length expressions, and those expressions are not effectively simplified. At the end of the computation, a matrix of $O(N^m)$-length expressions is obtained if $m+1$ matrices of size $N \times  N$ are multiplied together, which takes exponential time to simplify. In our approach, we represent matrices symbolically and simplify intermediate expressions efficiently on the fly, which has a much better performance.

	\section{Related work} \label{sec:related}
	
	Formal verification in quantum computing has been growing rapidly, especially in Coq.
	Boender et al.~\cite{BKN15} presented a framework for modeling and analyzing quantum protocols using Coq. They made use of the Coq repository C-CoRN~\cite{LCHF04} and built a matrix library with dependent types.
	Cano et al.~\cite{CCDMS16} specifically designed CoqEAL, a library built on top of ssreflect~\cite{ssref} to develop efficient computer algebra programs with proofs of correctness. They represented a matrix as a list of lists for efficient generic matrix computation in Coq but they did not consider optimizations specific for matrices commonly used in quantum computation.
	Paykin et al.~\cite{PRZ17} defined a quantum circuit language QWIRE in Coq, and formally verified some  quantum programs expressed in that language~\cite{RPZ18,RPLZ19}. 
	Reasoning using their matrix library usually requires explicit computation, which does not scale well, as discussed in Section~\ref{sec:exp}.
	Hietala et al.~\cite{HRHWH19} developed a quantum circuit compiler VOQC in Coq, which uses several peephole optimization techniques such as replacement, propagation, and cancellation as proposed by Nam et al.~\cite{YNYA18} to reduce the number of unitary transformations. It is very different from our symbolic approach of simplifying matrix operations using the Dirac notation. 
	Mahmoud et al.~\cite{MF19} formalized the semantics of Proto-Quipper in Coq and formally proved the type soundness property. They developed a linear logical framework within the Hybrid system~\cite{FM12}
	and used it to represent and reason about the linear type system of Quipper~\cite{GLRSV13}.  
	
	Our formalization of matrices is partly based on QWIRE~\cite{PRZ17}. We choose to use their formalization of matrices and matrix operations ($\sdot$, $\times$, $+$, $\otimes$, $\dag$), but without well-formness assumptions.
	And apart from that, our formalization of gates using Dirac notation, the notion of observational equivalence for circuits, and our systematic tactic library are novel.
	
	Note that although sparse matrix computation is well studied in other areas of Computer Science, we are not aware of any library in Coq dedicated to sparse matrices. We consider the symbolic approach proposed in the current work as a contribution in this perspective. For example, let us consider the gate CX. It is represented as a sparse $4\times 4$ matrix with $4$ out of the $16$ entries being non-zero  (cf. Section~\ref{sec:bqm}). As we can see in (\ref{eq:cx}), its symblic representation is
	\begin{equation}\label{eq:cxb}
		B_0 \otimes I_2 + B_3\otimes (B_1+B_2),
	\end{equation}
	using the terms $I_2$ and $B_j$ ($j\in \{0,...,3\}$). Those terms are the basic building blocks of our formalization of quantum circuits and our symbolic reasoning. The expression in (\ref{eq:cxb})  can be viewed as a compact way of representing the sparse matrix of gate $CX$.
	Furthermore, representing sparse matrices using Dirac notation is convenient for readability and cancelling zero matrices due to orthogonality of basic vectors.


	Apart from Coq, other proof assistants have also been used to verify quantum circuits and programs. 
	Liu et al.~\cite{LZWYLLYZ19} used the theorem prover Isabelle/HOL~\cite{NPW02} to formalize a quantum Hoare logic~\cite{Yin16} and verify its soundness and completeness for partial correctness. Unruh [6] developed a relational quantum Hoare logic and implemented an Isabelle-based tool to prove the security of post-quantum cryptography and quantum protocols.
	Beillahi et al.~\cite{BMT19} verified quantum circuits with up to 190 two-qubit gates in HOL Light. It relies on the formalization of Hilbert spaces in HOL Light proposed by Mahmoud et al. in~\cite{MAT13}, where a number of laws about complex functions and linear operators are proved. Although linear operators correspond to matrices in the finite-dimension case, our results are not implied by those in ~\cite{BMT19,MAT13}. For instance, a quantum state in~\cite{BMT19} is necessarily a vector, which means that only pure states can be represented. In contrast, we can also deal with mixed states in the form of density matrices.
	Chareton et al.~\cite{Chr21} proposed a verification framework QBRICKS embedded in the Why3 deductive verification tool~\cite{WHY}. It represents quantum circuits by a variant of path-sum representation~\cite{Amy18} called parametrized path-sums and can describe circuits in a functional language efficiently. However, this representation is measurement-free,  which is different from the our approach as we can directly describe quantum measurements on quantum states represented by density matrices.  
	
	Notice that the laws in Table~\ref{t:core} play an important role in our symbolic reasoning of quantum circuits. Although they resemble to some laws in a ring, the matrices under our consideration can be of various dimensions and they do not form a ring. It is also critical that the multiplication of two matrices, e.g. a row vector and a column vector, could be a scalar number (and even zero). Thus, rings are not enough here. The proof-by-reflection technique for rings might be useful but are usually hard to develop. We have shown that the tactic-based method is already efficient in our application scenario, and also flexible for both fully-automated and interactive proofs.
	
	\section{Conclusion and future work}\label{sec:concl}
We have proposed a symbolic approach to reasoning about quantum circuits in Coq. 
It is based on a small set of equational laws which are exploited to design some simplification strategies. According to our case studies, the approach
is more efficient than the usual one of explicitly representing matrices and is well suited to be automated in Coq.

Notice that in the current work the Dirac notation  is restricted to states on a computational basis. However, the notation is general enough to express states on other bases as well as linear operators. We leave it as a future work to investigate the more general scenarios so as to take full advantage of the Dirac notation. 

Dealing with quantum circuits is our intermediate goal. More interesting algorithms such as the Shor's algorithm~\cite{Sho94} also require classical computation.
In the near future, we plan to formalize in Coq the semantics of a quantum programming language with both classical and quantum features.

\end{document}